\documentclass{amsart}
\usepackage{amsmath,amsthm}
\usepackage{amsfonts,amssymb}

\usepackage{rotating}
\usepackage{graphicx}
\usepackage{longtable}

\usepackage{enumerate}

\newtheorem{theorem}{Theorem}[section]
\newtheorem{corollary}[theorem]{Corollary}
\newtheorem{lemma}[theorem]{Lemma}
\newtheorem{proposition}[theorem]{Proposition}
\newtheorem{example}[theorem]{Example}

\newtheorem{definition}[theorem]{Definition}
\newtheorem{remark}[theorem]{Remark}

\theoremstyle{remark}

\newcommand{\nd}{{\rm and }}

\newcommand{\mR}{\mathbb{R}}
\newcommand{\mC}{\mathbb{C}}
\newcommand{\mN}{\mathbb{N}}
\newcommand{\mE}{\mathbb{E}}
\newcommand{\mZ}{\mathbb{Z}}
\newcommand{\mS}{\mathbb{S}}
\newcommand{\mP}{\mathbb{P}}
\newcommand{\mI}{\mathbb{I}}

\newcommand{\cD}{\mathcal{D}}

\newcommand{\cH}{\mathcal{H}}

\newcommand{\cF}{\mathcal{F}}
\newcommand{\cP}{\mathcal{P}}

\newcommand{\cS}{\mathcal{S}}
\newcommand{\cG}{\mathcal{G}}

\newcommand{\ux}{\underline{x}}
\newcommand{\uxb}{\underline{x} \grave{}}

\newcommand{\uyb}{\underline{y} \grave{}}
\newcommand{\uy}{\underline{y}}

\newcommand{\uA}{\underline{\alpha}}

\newcommand{\pjb}{\partial_{{x \grave{}}_{j}}}

\newcommand{\pI}{\partial_{x_i}}

\newcommand{\px}{\partial_x}
\newcommand{\py}{\partial_y}

\begin{document}
 
\title[Hermite polynomials in superspace]
{Orthogonality of Hermite polynomials in superspace and Mehler type formulae}

\author{Kevin Coulembier}
\address{Department of Mathematical Analysis\\
Ghent University\\ Krijgslaan 281, 9000 Gent\\ Belgium.}
\email{Coulembier@cage.ugent.be}
\author{Hendrik De Bie}
\address{Department of Mathematical Analysis\\
Ghent University\\ Krijgslaan 281, 9000 Gent\\ Belgium.}
\email{Hendrik.DeBie@UGent.be}
\author{Frank Sommen}
\address{Department of Mathematical Analysis\\
Ghent University\\ Krijgslaan 281, 9000 Gent\\ Belgium.}
\email{fs@cage.ugent.be}

\date{\today}
\thanks{K. Coulembier is as Ph.D. Fellow of the Research Foundation - Flanders (FWO). H. De Bie is a Postdoctoral Fellow of the Research Foundation - Flanders (FWO)}

\begin{abstract}
In this paper, Hermite polynomials related to quantum systems with orthogonal $O(m)$-symmetry, finite reflection group symmetry $\cG < O(m)$, symplectic symmetry $Sp(2n)$ and superspace symmetry $O(m) \times Sp(2n)$ are considered. After an overview of the results for $O(m)$ and $\cG$, the orthogonality of the Hermite polynomials related to $Sp(2n)$ is obtained with respect to the Berezin integral. As a consequence, an extension of the Mehler formula for the classical Hermite polynomials to Grassmann algebras is proven. Next, Hermite polynomials in a full superspace with $O(m) \times Sp(2n)$ symmetry are considered. It is shown that they are not orthogonal with respect to the canonically defined inner product. However, a new inner product is introduced which behaves correctly with respect to the structure of harmonic polynomials on superspace. This inner product allows to restore the orthogonality of the Hermite polynomials and also restores the hermiticity of a class of Schr
 \"odinger operators in superspace. Subsequently, a Mehler formula for the full superspace is obtained, thus yielding an eigenfunction decomposition of the super Fourier transform. Finally, the new results for the $Sp(2n)$- and $O(m)\times Sp(2n)$-symmetry are compared with the results in the different types of symmetry.
\end{abstract}

\maketitle

\tableofcontents

%%%%%%%%%%%%%%%%%%%%%%%%%%%%%%%%%%%%%%%%%%%%%%%%%%%%%%%%%
%Start of paper

\section{Introduction}

This paper is concerned with the development of harmonic analysis and the related quantum mechanics in superspace (see e.g. \cite{DBS5,DBE1,DBS9,CDBS1}) and more specifically the Hermite polynomials introduced in that setting (see \cite{DBS3}). 

To fix ideas, let us first consider the quantum harmonic oscillator in $\mR^{m}$ with rotational $O(m)$-symmetry. This system is described by the Schr\"odinger equation
\begin{equation}
-\frac{\nabla^{2}}{2} \psi + \frac{r^{2}}{2} \psi = E \psi
\label{HOclass}
\end{equation}
and is typically solved in two different ways: either one uses cartesian co-ordinates or one uses spherical co-ordinates, yielding two types of Hermite polynomials. The use of cartesian co-ordinates reduces the problem to $m$ one-dimensional oscillators and yields solutions of the type
\begin{equation}
\psi_{k_{1}, \ldots, k_{m}}^{b} \propto H_{k_{1}}(x_{1}) \ldots H_{k_{m}}(x_{m}) e^{-r^{2}/2} , \qquad E= \frac{m}{2} + \sum_{i=1}^{m} k_{i}
\label{cartHerm}
\end{equation}
with $H_{k}(x) = (-1)^{k} \exp{(-x^2)} \frac{d^k}{dx^k} \exp{(-x^2)} $ the one-dimensional Hermite polynomials. If, on the other hand, one uses spherical co-ordinates, the eigenfunctions are expressed in terms of spherical harmonics, yielding
\begin{equation}
\phi_{j,k,l}^{b} \propto  L_{j}^{\frac{m}{2} + k-1}(r^2) H_{k}^{(l)} e^{-r^{2}/2}  , \qquad E= \frac{m}{2} + (2j+ k)
\label{sphHerm}
\end{equation}
with $L_{\alpha}^{\beta}$ the generalized Laguerre polynomials and $H_{k}^{(l)}$ a basis for the space of spherical harmonics of degree $k$. We will adopt the name spherical Hermite polynomials in this case. Both techniques completely solve the quantum harmonic oscillator and yield bases $\{\psi_{k_{1}, \ldots, k_{m}}^{b}\}$ and $\{ \phi_{j,k,l}^{b} \}$ of $L_{2}(\mR^{m}, dV(\ux))$ which are moreover orthonormal.

Second, it is possible to restrict the $O(m)$-symmetry in equation (\ref{HOclass}) to a finite reflection group $\cG < O(m)$. The related quantum system is then of Calogero-Moser-Sutherland type (see \cite{vD}) and given by
\[
-\frac{\Delta_{\kappa}}{2} \psi + \frac{r^{2}}{2} \psi = E \psi
\]
where $\Delta_{\kappa}$ is the so-called Dunkl Laplacian related to $\cG$ (see e.g. \cite{MR951883}, \cite{MR1827871}). Again, this equation can be solved using two types of Hermite polynomials. The first type is a generalization of the cartesian type (formula (\ref{cartHerm})) and was introduced by R\"osler (see \cite{MR1620515}). The second type (see \cite{MR1199124}) generalizes the spherical Hermite polynomials (formula (\ref{sphHerm})) to the Dunkl setting. (Explicit formulae will be presented in section \ref{DunklHermite}).
Again both types of functions form orthonormal bases for the weighted $L_{2}$-space $L_{2}(\mR^{m}, w_{\kappa}(\ux) dV(\ux))$. In this notation, $w_{\kappa}(\ux) dV(\ux)$ is the $\cG$-invariant measure in $\mR^{m}$.

Let us now turn our attention to the problem at hand. Again we make a change of symmetry by considering the symplectic group $Sp(2n)$ instead of $O(m)$ or $\cG$. The corresponding quantum problem is now formulated in a Grassmann algebra (which can be seen as a purely fermionic superspace). Again, as we will show in section \ref{FermionicMehler} and the beginning of section \ref{supersuper}, two bases of this Grassmann algebra exist (mimicking the previously discussed bases), which are now orthogonal with respect to a canonically defined inner product using the Berezin integral (see \cite{MR732126}).

However, the picture changes dramatically when one considers a full superspace with symmetry $O(m)\times Sp(2n)$. Although there still exist two types of Hermite polynomials, we will prove that the spherical Hermite polynomials are in this case not orthogonal with respect to the canonically defined inner product. As a consequence, Schr\"odinger operators for e.g. anharmonic oscillators in superspace are not self-adjoint with respect to this inner product, and we also do not immediately have an $O(m)\times Sp(2n)$ invariant Mehler formula for the spherical Hermite polynomials. Recall that the classical Mehler formula (see e.g. \cite{Sz}) is given by
\begin{equation*}
\sum_{k=0}^{\infty} \frac{e^{i k \alpha}}{2^k k! \sqrt{\pi}} H_k(x) H_k(y) = \left(\pi (1- e^{2i \alpha})\right)^{-1/2} \exp{\frac{ 2 e^{i \alpha} x y - e^{2i \alpha}(-x^2 + y^2)}{1- e^{2i \alpha}}}
\end{equation*}
and connects the one-dimensional Hermite polynomials with the kernel of the (fractional) Fourier transform.

Let us thus summarize the main aims of this paper in three questions:
\begin{itemize}
\item \textbf{Q1:} Can we construct a new inner product in full superspace such that the spherical Hermite polynomials are orthogonal?
\item \textbf{Q2:} Can we restore the self-adjointness of a class of Schr\"odinger operators of anharmonic type?
\item \textbf{Q3:} Can we obtain an $O(m)\times Sp(2n)$ invariant Mehler formula?
\end{itemize}
We will provide a positive answer to these 3 questions. Because it is possible to split integration in superspace in radial and spherical parts, the main technical difficulty lies in finding a positive definite inner product on the space of (super) spherical harmonics. This is the subject of the technical lemmas \ref{SSin1}, \ref{SSin2}, \ref{SSin3} and \ref{SSin4}, where use is made of the decomposition of harmonics in superspace under the action of $O(m)\times Sp(2n)$ obtained in \cite{DBE1} and of recent results on integration over the supersphere (see \cite{CDBS1}). These lemmas culminate in theorem \ref{defsuper}, where the new inner product is given.

The paper is organized as follows. In section \ref{classHarmAnalysis} we give a brief review of what is known in both the case of $O(m)$ and $\cG < O(m)$ symmetry. We focus on the $\mathfrak{sl}_{2}$ algebra generated by the Laplace operator and the squared length of a vector, introduce the two types of Hermite polynomials and discuss their orthogonality. We show how they give rise to $O(m)$ and $\cG < O(m)$ invariant Mehler formulae. In section \ref{Hamaninsup} we give the basic notions on superspaces needed for the sequel. We start with introducing Grassmann algebras, then proceed to full superspaces and discuss the notion of Schr\"odinger equations in superspaces. Next, in section \ref{FermionicMehler} we construct an inner product on the Grassmann algebra and obtain its basic properties. We show that the  spherical Hermite functions, related to the symplectic symmetry, form an orthogonal basis of the Grassmann algebra with respect to this inner product. Finally we obtain 
 a
  Mehler formula for the kernel of the purely fermionic Fourier transform. In doing so, we also determine the reproducing kernel for spaces of fermionic harmonics and express them in terms of regularized Gegenbauer polynomials. In section \ref{supersuper} we first discuss in detail where the orthogonality of the spherical Hermite polynomials in a full superspace fails. We then proceed to construct a new inner product and show that this inner product satisfies the desired properties, thus solving questions \textbf{Q1} and \textbf{Q2}. We then discuss how the spherical Hermite polynomials give rise to an $O(m)\times Sp(2n)$ invariant Mehler formula, answering question \textbf{Q3}. In section \ref{conclSummary} we summarize our results. We present in two extensive tables the differences and analogies that exist between the different types of symmetries considered in this paper. This also serves as a list of notations. Finally, we give several directions for further research. In 
 the appendix we recall some well-known facts about Hermite, Laguerre and Gegenbauer polynomials on the real line that will be used implicitly throughout the paper.

\section{Orthogonality of Hermite polynomials for $O(m)$ and finite reflection groups}
\setcounter{equation}{0}
\label{classHarmAnalysis}

\subsection{Classical harmonic analysis}
\label{ClassHO}

Harmonic analysis in $\mR^{m}$ is governed by the following three operators
\begin{eqnarray*}
\nabla^{2}_b &=& \sum_{i=1}^{m}\partial_{x_{i}}^{2}\\
r^{2} &=&  \sum_{i=1}^{m}x_{i}^{2}\\
\mE_{b} &=& \sum_{i=1}^{m} x_{i} \partial_{x_{i}}
\end{eqnarray*}
with $\Delta_{b}$ the Laplace operator and $\mE_{b}$ the Euler operator. The subindex $b$ denotes that we are working with bosonic or commuting co-ordinates. The operators $E = r^{2}/2$, $F =-\nabla^2_{b}/2$ and $H =\mE_{b} + m/2$ are invariant under $O(m)$ and generate the Lie algebra $\mathfrak{sl}_{2}$ (see e.g. \cite{MR1151617}):
\begin{equation}
\label{sl2relclass}
\big[H,E\big] = 2E,\>\> \big[H,F\big] = -2F,\>\>\big[E,F\big] = H.
\end{equation}

The space of polynomials in $\mR^{m}$ is given by $Pol = \mR[x_{1}, \ldots, x_{m}]$ and the space of homogeneous polynomials of degree $k$ by $Pol_{k}$. We then define the space $\cH_{k}^{b}$ of spherical harmonics of degree $k$ by $\cH_{k}^{b} = \ker{\nabla^2_{b}} \cap Pol_{k}$.

The Schr\"odinger equation of the harmonic oscillator is given by the following partial differential equation
\begin{equation}
-\frac{\nabla^2_{b}}{2} \psi + \frac{r^{2}}{2} \psi = E \psi
\end{equation}
and has two complete sets of solutions. Using cartesian co-ordinates one obtains
\begin{equation}
\psi_{k_{1}, \ldots, k_{m}}^{b} = \frac{1}{\sqrt{2^{\sum_{i=1}^{m} k_{i}} k_{1}! \ldots k_{m}! \pi^{m/2}}} H_{k_{1}}(x_{1}) \ldots H_{k_{m}}(x_{m}) e^{-r^{2}/2}
\label{cartbasisorthogonal}
\end{equation}
with $k_{i} \in \mN$ and with $H_{k}(x)$ the one-dimensional Hermite polynomials. The energy associated to $\psi_{k_{1}, \ldots, k_{m}}^{b}$ is given by $E= \frac{m}{2} + \sum_{i=1}^{m} k_{i}$. 

If one uses spherical co-ordinates, the eigenfunctions are expressed in terms of spherical harmonics, yielding the so-called spherical Hermite functions (see e.g. \cite{MR926831})
\begin{equation}
\phi_{j,k,l}^{b} =  \frac{1}{\sqrt{\frac{1}{2}4^{2j}j!\Gamma (j+\frac{m}{2}+k)}} \left[ (-\nabla^2_{b} - 4 r^2 + 4\mE_{b} + 2m)^{j} H_{k}^{(l)} \right] e^{-r^{2}/2},
\label{CHbosbasis}
\end{equation}
where the associated energy is $E= \frac{m}{2} + (2j+ k)$ and with $j, k \in \mN$. In (\ref{CHbosbasis}), $\{  H_k^{(l)}  \}$ ($l \in 1, \ldots, \dim \cH_{k}^{b}$) denotes a (real) orthonormal basis of $\cH_{k}^{b}$, i.e.
\begin{equation}
\label{bosharmbasis}
\int_{\mS^{m-1}} H_k^{(l_{1})}(\xi) \overline{H_k^{(l_{2})}}(\xi) d \sigma(\xi) = \delta_{l_{1} l_{2}},
\end{equation}
with $d\sigma$ the unique $O(m)$-invariant measure on $\mS^{m-1}$.

These functions can be written more explicitly as
\begin{eqnarray*}
\phi_{j,k,l}^{b}&=& \sqrt{\frac{2 j!}{\Gamma (j+\frac{m}{2}+k)}}   L_{j}^{\frac{m}{2} + k-1}(r^2) H_k^{(l)}e^{-r^{2}/2}
\end{eqnarray*}
with $L_{j}^{\frac{m}{2} + k-1}$ the generalized Laguerre polynomials.

Both the sets $\{\psi_{k_{1}, \ldots, k_{m}}^{b}\}$ and $\{ \phi_{j,k,l}^{b} \}$ are orthonormal bases of $L_{2}(\mR^{m}, dV(\ux))$:
\begin{eqnarray*}
\langle \psi_{k_{1}, \ldots, k_{m}}^{b} , \psi_{l_{1}, \ldots, l_{m}}^{b} \rangle_{L_{2}} = \int_{\mR^{m}} \psi_{k_{1}, \ldots, k_{m}}^{b} \overline{\psi_{l_{1}, \ldots, l_{m}}^{b}} dV(\ux) &=& \delta_{k_{1} l_{1}} \ldots \delta_{k_{m} l_{m}}\\
\langle \phi_{j_{1},k_{1},l_{1}}^{b} , \phi_{j_{2},k_{2},l_{2}}^{b} \rangle_{L_{2}}=\int_{\mR^{m}} \phi_{j_{1},k_{1},l_{1}}^{b} \overline{\phi_{j_{2},k_{2},l_{2}}^{b}} dV(\ux) &=& \delta_{j_{1} j_{2}} \delta_{k_{1} k_{2}} \delta_{l_{1} l_{2}},
\end{eqnarray*}
with $dV(\ux)$ the Lebesgue measure in $\mR^{m}$.

Recall that the classical Fourier transform is given by
\begin{equation}
\cF^{-}_{m|0}(f) = (2 \pi)^{-\frac{m}{2}} \int_{\mR^{m}} e^{-i\langle \ux,\uy\rangle} f(\ux) dV(\ux), \quad \langle \ux,\uy\rangle = \sum_{i=1}^{m} x_{i}y_{i}
\label{classFT}
\end{equation}
or in exponential operator notation by
\[
\cF^{-}_{m|0} = e^{ \frac{i \pi m}{4}} e^{\frac{i \pi}{4}(\nabla^2_{b} - r^{2})}.
\]
Again, both sets $\{\psi_{k_{1}, \ldots, k_{m}}^{b}\}$ and $\{ \phi_{j,k,l}^{b} \}$ act as eigenfunction bases for the Fourier transform
\begin{eqnarray*}
\cF^{-}_{m|0}(\psi_{k_{1}, \ldots, k_{m}}^{b}) &=& (-i)^{\sum_{i=1}^{m}k_{i}}\psi_{k_{1}, \ldots, k_{m}}^{b}\\
\cF^{-}_{m|0}(\phi_{j,k,l}^{b}) &=&(-i)^{2j+k}\phi_{j,k,l}^{b}.
\end{eqnarray*}

The Mehler formula for the one-dimensional Hermite polynomials is given by
\begin{equation*}
\sum_{k=0}^{\infty} \frac{e^{i k \alpha}}{2^k k! \sqrt{\pi}} H_k(x) H_k(y) = \left(\pi (1- e^{2i \alpha})\right)^{-1/2} \exp{\frac{ 2 e^{i \alpha} x y - e^{2i \alpha}(x^2 + y^2)}{1- e^{2i \alpha}}}.
\end{equation*}
If $\alpha = - \pi/2$ this formula gives a decomposition of the kernel of the one-dimensional Fourier transform. For a discussion of this formula, we refer the reader to \cite{Sz} or \cite{W}. A nice combinatorial proof can be found in \cite{MR0498167}.

There exist several generalizations of this formula. We will discuss the $O(m)$- and $\cG$-invariant cases in this and the next section. Note that there also exists a Mehler formula for the $q$-Hermite polynomials, see e.g. \cite{MR578207}.

The one-dimensional Mehler formula has an important property that can be deduced from the proof of the Mehler formula in \cite{W}.

\begin{lemma}
\label{Mehler1}
For any polynomial $D$ in four variables, one has
\begin{eqnarray*}
&&\sum_{k=0}^{\infty}D(x,y,\px,\py) \frac{e^{i k \alpha}}{2^k k! \sqrt{\pi}} H_k(x) H_k(y)\\ &=& \left(\pi (1- e^{2i \alpha})\right)^{-1/2} D(x,y,\px,\py)\exp{\frac{ 2 e^{i \alpha} x y - e^{2i \alpha}(x^2 + y^2)}{1- e^{2i \alpha}}}.
\end{eqnarray*}
The Hermite polynomials are defined as above and the series is absolutely convergent for $x,y,\alpha\in\mR$.
\end{lemma}

One can also construct $m$-dimensional Mehler formulae. Multiplying $m$ copies of the one-dimensional Mehler formula yields 
\begin{eqnarray*}
&&\sum_{k_{1}, \ldots, k_{m}} e^{i \alpha \sum_{i=1}^{m} k_{i}}\psi_{k_{1}, \ldots, k_{m}}^{b}(\ux) \psi_{k_{1}, \ldots, k_{m}}^{b}(\uy)\\ &=&  \left(\pi (1- e^{2i \alpha})\right)^{-\frac{m}{2}} e^{\frac{ 4 e^{i \alpha}\langle \ux,\uy \rangle - (1 +e^{2i \alpha})(r^2 + r_{\uy}^2)}{2- 2e^{2i \alpha}}}, \qquad  r_{\uy}^2 = \sum_{i=1}^{m} y_{i}^{2}.
\end{eqnarray*}
As this series is absolutely convergent, we can rearrange terms in the left-hand side. Using the fact that the change of basis from $\{\psi_{k_{1}, \ldots, k_{m}}^{b}\}$ to $\{ \phi_{j,k,l}^{b} \}$ is orthogonal in each eigenspace of the harmonic oscillator we obtain
\[
\sum_{j,k,l} e^{i \alpha (2j+k)}\phi_{j,k,l}^{b}(\ux) \phi_{j,k,l}^{b}(\uy) =  \left(\pi (1- e^{2i \alpha})\right)^{-\frac{m}{2}} e^{\frac{ 4 e^{i \alpha}\langle \ux,\uy \rangle - (1 +e^{2i \alpha})(r^2 + r_{\uy}^2)}{2- 2e^{2i \alpha}}}.
\]
This formula can be simplified even further by making use of the reproducing kernel of the spaces of spherical harmonics, given by (see e.g. \cite{MR0499342})
\begin{align}
\label{bosreprkern}
\begin{split}
F_k(\ux,\uy)  &=\sum_{l=1}^{\dim \cH_{k}^{b}}H_{k}^{(l)}(\ux) H_{k}^{(l)}(\uy)\\ &=  \frac{2k+m-2}{m-2} \frac{\Gamma(m/2)}{2 \pi^{m/2}} (|\ux||\uy|)^k  C^{(m-2)/2}_k (\langle \frac{\ux}{|\ux|}, \frac{\uy}{|\uy|}\rangle) 
\end{split}
\end{align}
with $C^{(m-2)/2}_k$ the Gegenbauer polynomial of degree $k$ (see Appendix) and where we have homogenized the right-hand side. This yields the $O(m)$-invariant version of the Mehler formula for every $\ux,\uy\in\mR^m$ and $\alpha\in\mR$:
\begin{align}
\label{OmMehler2}
\begin{split}
&\sum_{j,k}\frac{2j! e^{i \alpha (2j+k)}}{\Gamma(j+\frac{m}{2}+k)} L_j^{\frac{m}{2}+k-1}(r^2)L_j^{\frac{m}{2}+k-1}(r_{\uy}^2)F_k(\ux,\uy) \\ 
=& \left(\pi (1- e^{2i \alpha})\right)^{-\frac{m}{2}} e^{\frac{  2 e^{i \alpha}\langle \ux,\uy \rangle - e^{2i \alpha}(r^2 + r_{\uy}^2)}{1- e^{2i \alpha}}}.
\end{split}
\end{align}
In this formula only $r^2$, $r_{\uy}^2$ and $\langle \ux,\uy \rangle$ appear, which are three arbitrary real numbers satisfying $r^2\ge 0$, $r_{\uy}^2\ge 0$ and $\langle \ux,\uy \rangle^2\le r^2r_{\uy}^2$. This leads to the following corollary.

\begin{corollary}
For every $a,b\in\mR^+$ and $c\in\mR$ with $c^2\le a^2b^2$, one has
\begin{eqnarray*}
&&\sum_{j,k}\frac{2j! e^{i \alpha (2j+k)}}{\Gamma(j+\frac{m}{2}+k)} L_j^{\frac{m}{2}+k-1}(b^2)L_j^{\frac{m}{2}+k-1}(a^2)(ab)^k G_k\left(\frac{c}{ab}\right)\\ 
&=& \left(\pi (1- e^{2i \alpha})\right)^{-m/2} \exp{\frac{  2 e^{i \alpha} c  - e^{2i \alpha}(a^2 +b^2)}{1- e^{2i \alpha}}}
\end{eqnarray*}
where 
\[
G_k\left(\frac{c}{ab}\right) = \frac{2k+m-2}{m-2} \frac{\Gamma(m/2)}{2 \pi^{m/2}} \,  C^{(m-2)/2}_k \left(\frac{c}{ab}\right). 
\]
\label{OmMehler3}
\end{corollary}
In this way, the Mehler formula has been stripped of its geometrical meaning in $\mR^{m}$ and is reduced to a statement about orthogonal polynomials on the real line. We will need this in the proof of the $O(m)\times Sp(2n)$-invariant Mehler formula. 

\begin{remark}
\label{Mehlerafleiden}
Lemma \ref{Mehler1} is immediately adapted to the $O(m)$-invariant case in equation (\ref{OmMehler2}) or corollary \ref{OmMehler3}.
\end{remark}

\subsection{Hermite polynomials related to the Dunkl Laplacian}
\label{DunklHermite}

Denote by $\langle .,. \rangle$ the standard Euclidean scalar product in $\mR^{m}$ and by $r = \langle \ux, \ux\rangle^{1/2}$ the associated norm. For $\uA \in \mR^{m} - \{ 0\}$, the reflection $r_{\uA}$ in the hyperplane orthogonal to $\uA$ is given by
\[
r_{\uA}(\ux) = \ux - 2 \frac{\langle \uA, \ux\rangle}{|\uA|^{2}}\uA, \quad \ux \in \mR^{m}.
\]

A root system is a finite subset $R \subset \mR^{m}$ of non-zero vectors such that, for every $\uA \in R$, the associated reflection $r_{\uA}$ preserves $R$. We will assume that $R$ is reduced, i.e. $R \cap \mR \uA = \{ \pm \uA\}$ for all $\uA \in R$. Each root system can be written as a disjoint union $R = R_{+} \cup (-R_{+})$, where $R_{+}$ and $-R_{+}$ are separated by a hyperplane through the origin. The subgroup $\cG \subset O(m)$ generated by the reflections $\{r_{\uA} | \uA \in R\}$ is called the finite reflection group associated with $R$. We will also assume that $R$ is normalized such that $\langle \uA, \uA\rangle = 2$ for all $\uA \in R$. For more information on finite reflection groups we refer the reader to \cite{Humph}.

A multiplicity function $\kappa$ on the root system $R$ is a $\cG$-invariant function $\kappa: R \rightarrow \mC$, i.e. $\kappa(\uA) = \kappa(h \uA)$ for all $h \in \cG$. We will denote $\kappa(\uA)$ by $\kappa_{\uA}$.

Fixing a positive subsystem $R_{+}$ of the root system $R$ and a multiplicity function $\kappa$, we introduce the Dunkl operators $T_{i}$ associated to $R_{+}$ and $\kappa$ by (see \cite{MR951883}, \cite{MR1827871})
\[
T_{i} f(\ux)= \partial_{x_{i}} f(\ux) + \sum_{\uA \in R_{+}} \kappa_{\uA} \alpha_{i} \frac{f(\ux) - f(r_{\uA}(\ux))}{\langle \uA, \ux\rangle}, \qquad f \in C^{1}(\mR^{m}).
\]
An important property of the Dunkl operators is that they commute, i.e. $T_{i} T_{j} = T_{j} T_{i}$.

The Dunkl Laplacian is given by $\Delta_{\kappa} = \sum_{i=1}^{m} T_i^2$, or more explicitly by
\[
\Delta_{\kappa} f(\ux) = \nabla^{2} f(\ux) + 2 \sum_{\uA \in R_{+}} \kappa_{\uA} \left( \frac{\langle \nabla f(\ux), \uA \rangle}{\langle \uA, \ux \rangle}  - \frac{f(\ux) - f(r_{\uA}(\ux))}{\langle \uA, \ux \rangle^{2}} \right)
\]
with $\nabla$ the gradient operator.

If we let $\Delta_{\kappa}$ act on $r^2$ we find $\Delta_{\kappa} r^2 = 2m + 4 \gamma = 2 \mu$, where $\gamma = \sum_{\uA \in R_+} \kappa_{\uA}$. We refer to $\mu$ as the Dunkl dimension, because most special functions related to $\Delta_{\kappa}$ behave as if one would be working with the classical Laplace operator in a space with dimension $\mu$. 

We also denote by $\cH_k^{\cD}$ the space of Dunkl-harmonics of degree $k$, i.e. $\cH_k^{\cD} = \ker{\Delta_{\kappa}} \cap Pol_k$. The space of Dunkl-harmonics of degree $k$ has the same dimension as the classical space of spherical harmonics of degree $k$ and a basis can e.g. be constructed using Maxwell's representation (see \cite{Xu}).

It is possible to construct an intertwining operator $V_{\kappa}$ connecting the classical derivatives $\partial_{x_{j}}$ with the Dunkl operators $T_{j}$ such that $T_{j} V_{\kappa} = V_{\kappa} \partial_{x_{j}}$ (see e.g. \cite{MR1273532}). Note that explicit formulae for $V_{\kappa}$ are only known in a few special cases.

The operators
\[ E:= \frac{1}{2}r^2,\>\> F:= -\frac{1}{2}\Delta_{\kappa} \quad\text{and}\>\> 
H:= \mE +\mu/2
\]
again satisfy the defining relations \eqref{sl2relclass} of the Lie algebra $\mathfrak{sl}_2$, see \cite{He}.

We further define the Fischer inner product on $Pol$ by (see \cite{MR1145585})
\[
[p,q]_{\kappa} = \left(p(T/2) q\right) (0), \qquad p,q \in Pol
\]
where $p(T)$ means substituting the variables $x_{i}$ for Dunkl operators $T_{i}$. Using an analog of a result of MacDonald (see \cite{MR0558859}), this inner product can be rewritten as
\begin{equation}
[p,q]_{\kappa} = d_{\kappa}^{-1} \int_{\mR^{m}} \left(e^{-\Delta_{\kappa}/4} p(\ux)\right)\left(e^{-\Delta_{\kappa}/4} q(\ux)\right) e^{-r^{2}} w_{\kappa}(\ux)dV(\ux), \label{FischerMcDo}
\end{equation}
with $w_{\kappa}(\ux) = \prod_{\alpha \in R_{+}} |\langle \uA, \ux\rangle |^{2 \kappa_{\alpha}}$ the weight function corresponding to $\cG$ and with $d_{\kappa} =\int_{\mR^{m}} e^{-r^{2}} w_{\kappa}(\ux)dV(\ux)$ (see e.g. \cite{MR2022853}).

Now let $\{p_\nu\,, \nu\in \mZ_+^m\}$ be a basis of $Pol$ such that $p_\nu\in \mathcal Pol_{|\nu|}$ (with $|\nu|=\sum_{i=1}^m\nu_i$) and moreover $[p_\nu, p_\mu]_{\kappa} = \delta_{\nu \mu}$. The `cartesian' Hermite polynomials related to $\cG$ are defined as follows by R\"osler (see \cite{MR1620515,MR2022853}).

\begin{definition}
\label{DefHermRos}
The generalized Hermite polynomials $\{H_\nu\,, \>\nu\in \mZ_+^m\}$  
associated with the basis $\{p_\nu\}$ on $\mR^m$ are given by 
\[
H_\nu(\ux):= 2^{|\nu|}e^{-\Delta_{\kappa}/4}p_\nu(\ux) = 
2^{|\nu|}\sum_{n=0}^{\lfloor|\nu|/2\rfloor} \frac{(-1)^n}{4^n n!}\,
\Delta_{\kappa}^n p_\nu(\ux).
\]
Moreover, the generalized Hermite functions on $\mR^m$ are then defined by
\[
\psi_\nu^{\cD}(\ux):= \frac{1}{2^{|\nu|} \sqrt{d_{\kappa} }}H_\nu(\ux) e^{-r^2/2}, \quad \nu\in \mZ_+^m.
\]
\end{definition}

\begin{remark}
Similar exponential formulas for Hermite polynomials have been given earlier in \cite{MR1471336} by Baker and Forrester, following (unpublished) work of Lassalle. The Hermite polynomials given in Definition \ref{DefHermRos} yield the polynomials of \cite{MR1471336} after symmetrization. 

When $\cG$ is the group $S_{m}$ or the group $B_{m}$, the polynomials in Definition \ref{DefHermRos} include, under a suitable choice of basis $\{p_\nu\}$, the polynomials studied in \cite{MR1456121, MR1646546} and previously introduced in symmetric version in \cite{MR1105634, MR1133488}.
\end{remark}

The Rodrigues formula for $H_{\nu}(\ux)$ is given by
\[
H_{\nu}(\ux) = (-1)^{|\nu|} e^{r^2}p_{\nu}(\underline{T})e^{-r^2},
\]
so if we consider $\kappa=0$ and take the standard monomial basis of $Pol$ we reobtain the cartesian Hermite polynomials for the orthogonal case.

Using formula (\ref{FischerMcDo}) we immediately obtain that the set $\{\psi_\nu^{\cD} \}$ forms an orthonormal basis of the weighted $L_{2}$-space $L_{2}(\mR^{m}, w_{\kappa}(\ux) dV(\ux))$ :
\[
\int_{\mR^{m}} \psi_{\nu}^{\cD} \overline{\psi_{\mu}^{\cD}} w_{\kappa}(\ux) dV(\ux) =  \delta_{\nu \mu}.
\]

The related quantum system in this case is given by the following PDE with difference terms
\begin{equation}
-\frac{\Delta_{\kappa}}{2} \psi + \frac{r^{2}}{2} \psi = E \psi,
\label{HOdunkl}
\end{equation}
which is a so-called Calogero-Moser-Sutherland system (with harmonic confinement). For a review of this type of quantum systems we refer the reader to \cite{vD,MR2022853}.
If $p_\nu$ is a homogeneous polynomial of degree $|\nu|$, then the associated Hermite function $\psi_{\nu}^{\cD}$ is a solution of (\ref{HOdunkl}) with corresponding energy $E = \mu/2 + |\nu|$ (see \cite{MR1620515}).

It is also possible to introduce a generalization of the spherical Hermite functions related to the Dunkl Laplacian (see a.o. \cite{MR1199124,Said} or \cite{DBSIGMA} for an approach using Clifford analysis). They are defined as follows
\begin{equation}
\phi_{j,k,l}^{\cD} =  \frac{1}{\sqrt{\frac{1}{2}4^{2j}j!\Gamma (j+\frac{\mu}{2}+k)}} \left[ (-\Delta_{\kappa} - 4 r^2 + 4\mE_{b} + 2\mu)^{j} H_{k}^{(l)} \right] e^{-r^{2}/2},
\end{equation}
with associated energy $E= \frac{\mu}{2} + (2j+ k)$ and where $j, k \in \mN$. In this notation, $\{  H_k^{(l)} \}$, $l \in 1, \ldots, \dim \cH_{k}^{\cD}$, is a (real) orthonormal basis of $\cH_{k}^{\cD}$ satisfying
\[
\int_{\mS^{m-1}} H_k^{(l_{1})}(\xi) \overline{H_k^{(l_{2})}}(\xi) w_{\kappa}(\xi) d \sigma(\xi) =  \delta_{l_{1} l_{2}}.
\]
More explicitly, they are given by
\begin{eqnarray*}
\phi_{j,k,l}^{\cD}&=& \sqrt{\frac{2 j!}{\Gamma (j+\frac{\mu}{2}+k)}}   L_{j}^{\frac{\mu}{2} + k-1}(r^2) H_k^{(l)}e^{-r^{2}/2}.
\end{eqnarray*}
The set of functions $\{ \phi_{j,k,l}^{\cD} \}$ again forms an orthonormal basis of the space $L_{2}(\mR^{m}, w_{\kappa}(\ux) dV(\ux))$, i.e.
\[
\int_{\mR^{m}} \phi_{j_{1},k_{1},l_{1}}^{\cD} \overline{\phi_{j_{2},k_{2},l_{2}}^{\cD}} w_{\kappa}(\ux) dV(\ux) = \delta_{j_{1} j_{2}} \delta_{k_{1} k_{2}} \delta_{l_{1} l_{2}}.
\]

There exists a Fourier transform related to the Dunkl Laplacian (see \cite{MR1199124,deJ}). This so-called Dunkl transform $\mathcal \cF_{\kappa}: L^1(\mR^m, w_{\kappa}(\ux)dx)\to C(\mR^m)$ is defined as follows
\[
 \mathcal \cF_{\kappa} f(\uy):= c_{\kappa}^{-1} \int_{\mR^m} f(\ux)\,D(\ux,-i\uy)\,w_{\kappa}(\ux) dV(\ux) \quad 
(\uy \in \mR^m)
\]
with $c_{\kappa} = \int_{\mR^m} e^{-r^{2}/2}w_{\kappa}(\ux) dV(\ux)$ the Mehta constant related to $\cG$ and where $D(\ux,\uy)$ is the Dunkl kernel. This kernel is the unique solution of the system
\[
T_{i, x} D(\ux,\uy) = y_{i} D(\ux,\uy), \quad i=1, \ldots, m
\]
which is real-analytic in $\mR^{m}$ and satisfies $K(0,\uy)=1$. For general reflection groups this kernel is not explicitly known. There also exists an operator exponential expression for the Dunkl transform
\begin{equation*}
 \cF_{\kappa} = e^{ \frac{i \pi \mu}{4}} e^{\frac{i \pi}{4}(\Delta_{\kappa} - r^{2})}
\end{equation*}
as has been studied in-depth in \cite{Said}. Note that both bases $\{\psi_\nu^{\cD} \}$ and $\{ \phi_{j,k,l}^{\cD} \}$ are eigenfunctions of the Dunkl transform, i.e.
\begin{eqnarray*}
 \cF_{\kappa} (\psi_\nu^{\cD}) &=& (-i)^{|\nu|} \psi_\nu^{\cD}\\
 \cF_{\kappa} (\phi_{j,k,l}^{\cD}) &=& (-i)^{2j + k} \phi_{j,k,l}^{\cD}.
\end{eqnarray*}
For the first basis this was obtained in  \cite{MR1620515}, for the second basis in \cite{MR1199124,Said}.

A Mehler formula for the `cartesian' Hermite polynomials related to $\cG$ was obtained in \cite{MR1620515}, Theorem 3.12. It states that, for $u\in \mC$ with $|u|<1$ and all $\ux, \uy \in \mR^{m}$, the following holds
\[
\sum_{\nu\in \mZ_+^m} d_{\kappa} u^{|\nu|} \psi_\nu^{\cD}(\ux) \psi_\nu^{\cD}(\uy) =(1-u^{2})^{-\mu/2} D\left( \frac{2u\ux}{1-u^{2}},\uy\right)e^{-\frac{1+u^{2}}{2-2u^{2}}(r^{2}+r_{\uy}^{2})}.
\]
Going over to the basis $\{\phi_{j,k,l}^{\cD}\}$ we then immediately obtain
\[
\sum_{j,k,l} d_{\kappa} u^{2j+k} \phi_{j,k,l}^{\cD}(\ux) \phi_{j,k,l}^{\cD}(\uy) =(1-u^{2})^{-\mu/2} D\left( \frac{2u\ux}{1-u^{2}},\uy\right)e^{-\frac{1+u^{2}}{2-2u^{2}}(r^{2}+r_{\uy}^{2})}.
\]
Note that when $\kappa=0$, then $d_{\kappa}= \pi^{m/2}$ and the formulae reduce to the formulae in section \ref{ClassHO}. The last formula can be simplified by using the explicit expression of $\{\phi_{j,k,l}^{\cD}\}$ in terms of generalized Laguerre polynomials and Dunkl harmonics. Then one can apply the reproducing kernel for $\cH_{k}^{\cD}$, which is again given by a suitable Gegenbauer polynomial, see \cite{MR1402890}, as
\begin{eqnarray*}
\sum_{l=1}^{\dim \cH_{k}^{\cD}}H_{k}^{(l)}(\ux/|\ux|) H_{k}^{(l)}(\uy/|\uy|) &=&  \frac{2k+\mu-2}{\mu-2} \frac{\Gamma(\mu/2)}{2^{(1-\mu)/2} c_{\kappa}}\\
&&\times [V_{\kappa} C^{(\mu-2)/2}_k \left(\langle \cdot , \uy/|\uy|\rangle \right)](\ux/ |\ux|), 
\end{eqnarray*}
with $V_{\kappa}$ the intertwining operator. Homogenizing this reproducing kernel then leads to a similar formula as (\ref{OmMehler2}), which is now $\cG$-invariant.

\begin{remark}
In \cite{MR2134314,MR2401813} and the recent preprint \cite{Orsted2}, the role played by Mehler type formulae for orthogonal and finite reflection group symmetries is further elucidated, for radially deformed operators still satisfying the defining relations of $\mathfrak{sl}_{2}$.
\end{remark}

\section{Harmonic analysis in superspace}
\setcounter{equation}{0}
\label{Hamaninsup}

\subsection{Grassmann algebras}

Consider the complex Grassmann algebra $\Lambda_{2n}$ generated by $2n$ anti-commuting generators ${x \grave{}}_i$ satisfying ${x \grave{}}_j {x \grave{}}_k = - {x \grave{}}_k {x \grave{}}_j$. An arbitrary element $f \in \Lambda_{2n}$ can always be written as $f = \sum_A f_A {x \grave{}}_A$ with ${x \grave{}}_A = {x \grave{}}_1^{\, \alpha_1} \ldots {x \grave{}}_{2n}^{\, \alpha_{2n}}$, $A = (\alpha_{1}, \ldots, \alpha_{2n}) \in \{0,1\}^{2n}$ and $f_A \in \mC$. The dimension of $\Lambda_{2n}$ as a $\mC$-vectorspace is hence $2^{2n}$. $\Lambda_{2n}$ decomposes as
\[
\Lambda_{2n} = \oplus_{k=0}^{2n} \Lambda_{2n}^{k}
\]
with $\Lambda_{2n}^{k}$ the space of homogeneous elements of degree $k$.

Introducing $\theta^2 = -\sum_{j=1}^n {x\grave{}}_{2j-1} {x\grave{}}_{2j}$ and the so-called fermionic Laplace operator $\nabla^2_f =- 4 \sum_{j=1}^n \partial_{{x \grave{}}_{2j-1}} \partial_{{x \grave{}}_{2j}}$, we have the following decomposition of $\Lambda_{2n}$ (see \cite{DBE1}):
\begin{equation}
\Lambda_{2n} = \bigoplus_{k=0}^{n} \left(\bigoplus_{j=0}^{n-k} \theta^{2j} \cH^f_k \right),
\label{Fischer}
\end{equation}
where $\cH^f_k$ is the space of fermionic spherical harmonics of degree $k$, i.e. $\cH_{k}^{f} = \ker \nabla^2_f \cap \Lambda_{2n}^{k}$. The dimension of $\cH^f_{k}$ is given by $\binom{2n}{k} - \binom{2n}{k-2}$. Formula (\ref{Fischer}) is the decomposition of $\Lambda_{{2n}}$ into irreducible pieces under the action of $Sp(2n)$ (see \cite{DBE1}).

The fermionic Euler operator is given by $\mE_f = \sum_{j=1}^{2n} {x \grave{}}_{j} \pjb$ and satisfies the equation
\begin{eqnarray}
\label{fermEuler}
[\nabla^2_f,\theta^2]&=&4\mE_f-4n.
\end{eqnarray}
We also have $[\nabla^2_{f},\mE_f-n ] = 2 \nabla^2_{f}$ and $[\theta^{2},\mE_f-n] = -2 \theta^{2}$ so the $\mathfrak{sl}_{2}$ relations from both the orthogonal and the Dunkl case extend to the Grassmann case. It is moreover easily seen that $\theta^2$, $-\nabla^2_f$ and $\mE_{f}-n$ are invariant under the action of $Sp(2n)$. So we have obtained a change of symmetry from either $O(m)$ or $\cG < O(m)$ to $Sp(2n)$.

\begin{remark}
In previous papers the notations $\Delta_f=-\nabla^2_f$ and $\uxb^2=-\theta^2$ were used.
\end{remark}

We have the following important lemma concerning the action of the Laplace operator (see \cite{DBS5}, Corollary 2). 
\begin{lemma}
\label{Laplfischer}
Let $H_k \in \cH^f_k$. Then for $j \leq n-k $ one has
\[
\nabla^{2i}_f(\theta^{2j} H_k) = \left\{
\begin{array}{l}
c_{i,j,k} \theta^{2j-2i} H_k, \quad i \leq j\\
\vspace{-2mm}\\
0, \quad i > j
\end{array}
\right.
\]
with
\[
c_{i,j,k} = (-1)^i 4^{i} \frac{j!}{(j-i)!} \frac{(n+i-j-k)!}{(n-j-k)!}.
\]
\end{lemma}

The fermionic Gaussian function is given by the finite Taylor expansion $\exp(-\theta^2/2) =\sum_{j=0}^n(-1)^j \theta^{2j}/(2^{j}j!)$.
We consider a basis $H_k^{(l)}$ of $\cH^f_k$. The functions
\begin{align}
\label{CliffordHermiteFunctions}
\begin{split}
\varphi_{j,k,l}^{f} =& (-\nabla^2_{f} -\theta^2 + 2\mE -2n)^j H_k^{(l)} \exp(-\theta^2/2)\\
 =&2^{2j}j! L_j^{-n+k-1}(\theta^2) H_k^{(l)}\exp(-\theta^2/2)
\end{split}
\end{align}
for $j = 0, \ldots, n-k$; $k = 0, \ldots, n$ and $l = 1, \ldots, \dim \cH_k^{f}$ and with $L_\beta^{\alpha}$ the generalized Laguerre polynomials, are the so-called fermionic spherical Hermite functions (see \cite{DBS3}). They constitute a basis of $\Lambda_{2n}$, which follows immediately from the decomposition (\ref{Fischer}). 

Note that it is possible to explicitly construct a basis of $\cH_k^f$ by decomposing this space under the action of the subgroup $Sp(2) \times Sp(2n-2)$ of $Sp(2n)$. This leads to the following theorem.
\begin{theorem}
\label{basisrecursief}
If $1<k\leq n$, then the space $\cH_k^f({x \grave{}}_1 , \ldots, {x \grave{}}_{2n})$ decomposes as
\begin{eqnarray*}
&& \cH_k^f({x \grave{}}_3 , \ldots, {x \grave{}}_{2n}) \;  \oplus \; \cH_1^f({x \grave{}}_1 ,{x \grave{}}_2)\otimes \cH_{k-1}^f({x \grave{}}_3 , \ldots, {x \grave{}}_{2n})\\
&& \oplus \;  \left[{x \grave{}}_1 {x \grave{}}_2
 + \frac{1}{k-n-1} ({x \grave{}}_3 {x \grave{}}_4 + \ldots +{x \grave{}}_{2n-1} {x \grave{}}_{2n}) \right] \cH_{k-2}^f({x \grave{}}_3 , \ldots, {x \grave{}}_{2n}).
\end{eqnarray*}
\end{theorem}

The integration we use on $\Lambda_{2n}$ is the so-called Berezin integral (see \cite{MR732126}, \cite{DBS5}), defined by
\[
\int_{B,x} = \pi^{-n} \partial_{{x \grave{}}_{2n}} \ldots \partial_{{x \grave{}}_{1}} = \frac{ \pi^{-n}}{4^n n!} \nabla_f^{2n}.
\]

\subsection{Full superspace}

Next we consider a general superspace, with $m$ commuting and $2n$ anti-commuting variables. The space of polynomials is then denoted by $\cP=\mR[x_1,\ldots,x_m]\otimes \Lambda_{2n}$. The full Laplace operator is given by
\[
\nabla^2=\nabla^2_b+\nabla^2_f = \sum_{i=1}^m\partial_{x_i}^2 -4 \sum_{j=1}^n \partial_{{x \grave{}}_{2j-1}} \partial_{{x \grave{}}_{2j}}
\]
and the generalization of $r^{2}$ is 
\[
R^2=r^2+\theta^2 = \sum_{i=1}^m x_i^2 - \sum_{j=1}^n {x\grave{}}_{2j-1} {x\grave{}}_{2j}.
\]
\begin{remark}
In previous papers the notations $\bold{x}^2=-R^2$ and $\Delta=-\nabla^2$ were used.
\end{remark}
The super Euler operator $\mE$ is given by $\mE = \mE_{b}+ \mE_{f}$. We can introduce a dimension parameter, the so-called super-dimension, by
\[
M=m-2n=\frac{1}{2}\nabla^2(R^2).
\]
The operators $E = R^{2}/2$, $F = -\nabla^2/2$ and $H =\mE + M/2$ again satisfy the defining commutation relations of $\mathfrak{sl}_{2}$ (see formula (\ref{sl2relclass})), although the symmetry of the operators is now given by $O(m)\times Sp(2n)$, see \cite{DBE1}. In particular
\begin{equation}
\label{commsuper}
\left[\nabla^2/2,R^2/2\right]=\mE+M/2
\end{equation}
holds. The symmetry group is realized as the matrices $A\in\mR^{(m+2n)\times(m+2n)}$ which satisfy $A^TGA=G$ with 
\begin{eqnarray}
\label{metricG}
G&=&\left( \begin{array}{c|c} \mI_m&0\\ \hline \vspace{-3.5mm} \\0&J
\end{array}\right)
\end{eqnarray}
and 
\[
J =  \left( \begin{array}{cccccc} 0&1&&&\\-1&0&&&\\&&\ddots&&\\&&&0&1\\&&&-1&0 
\end{array}
 \right).
\]
The action of $O(m)\times Sp(2n)$ on superfunctions is then given by
\begin{eqnarray}
\label{actieOSp}
Af(\bold{x})&=&f(A^T\bold{x}).
\end{eqnarray}

The space $\cH_{k}$ of (super) spherical harmonics of degree $k$ is defined by $\cH_{k} = \ker \nabla^2 \cap \cP_{k}$, with $\cP_{k}$ the space of homogeneous polynomials of degree $k$. If the super-dimension $M=m-2n$ is not even and negative, we have the following decomposition of $\cP$
\begin{equation}
\label{superFischer}
\cP = \bigoplus_{k=0}^{\infty} \cP_k= \bigoplus_{j=0}^{\infty} \bigoplus_{k=0}^{\infty} R^{2j}\cH_k.
\end{equation}
This decomposition is closely related to the Howe dual pair $(\mathfrak{osp}(m|2n),\mathfrak{sl}_2)$. Each subspace $\bigoplus_{j=0}^{\infty}  R^{2j}\cH_k$ is an irreducible $\mathfrak{sl}_2$-representation, the weight vectors are $R^{2j}\cH_k$. The blocks $R^{2j}\cH_k$ are exactly the irreducible pieces of the representation of $\mathfrak{osp}(m|2n)$ on $\cP$ when $M>0$, see \cite{ZHANG}.

There exist explicit projection operators for the decompositions (\ref{Fischer}) and (\ref{superFischer}), satisfying
\[
\mP_i^k (R^{2j} \cH_{k-2j}) = \delta_{ij} \cH_{k-2j}.
\]
They were constructed in \cite{DBE1} (see also \cite{DBThesis}). More specifically, $\mP_0^k$ is given by
\begin{eqnarray}
\label{projectie}
\mP_0^k &=& \sum_{j=0}^{\lfloor k/2 \rfloor} \frac{1}{4^{j} j!} \frac{1}{(k+M/2-j-1)_j}  R^{2j}\nabla^{2j}, 
\end{eqnarray}
with $(a)_j=a(a+1)\cdots(a+j-1)$ the Pochhammer symbol.

We can again introduce the spherical Hermite functions, and they will now form a basis of $\cP \exp (-r^2/2)$. Because $\exp(-\theta^2/2)$ is an invertible element of $\Lambda_{2n}$, we have 
\[\cP \exp (-r^2/2)=\cP\exp(-r^2/2)\exp(-\theta^2/2)=\cP \exp (-R^2/2).\] 
If we now consider a basis $H_k^{(l)}$ of $\cH_k$, then the functions
\begin{eqnarray*}
\varphi_{j,k,l}& =& (-\nabla^2 -R^2 + 2\mE +M)^j H_k^{(l)} \exp{(-R^2/2)} 
\label{CliffordHermiteFunctionssuper}
\end{eqnarray*}
with $j, k \in \mN$ and $l = 1, \ldots, \dim \cH_k$, are the so-called spherical Hermite functions in (full) superspace (see \cite{DBS3}). They constitute a basis of $\cP \exp (-R^2/2)$ when $M\not\in-2\mN$, which follows immediately from the decomposition (\ref{superFischer}). In terms of classical orthogonal polynomials the functions $\varphi_{j,k,l}$ are given by
\begin{equation}
\varphi_{j,k,l} = 2^{2j} j! L^{\frac{M}{2}+k-1}_{j}(R^2)   H_k^{(l)} \exp{(-R^2/2)}
\label{CHLaguerre}
\end{equation}
with $L^{\alpha}_{\beta}$ the generalized Laguerre polynomials (see \cite{DBS3}, theorem 8) when $M\not\in-2\mN$ or when $m=0$. In the limit $n\to 0$, the spherical Hermite functions (\ref{CliffordHermiteFunctionssuper}) reduce to the purely bosonic spherical Hermite functions (see \cite{MR926831} or formula (\ref{CHbosbasis})). The spherical Hermite functions satisfy
\begin{equation}
\label{differentiaalvgl}
(\nabla^2+R^2+2\mE+M)\varphi_{j,k,l}=-8j(2j+M+2k-2)\varphi_{j-1,k,l}
\end{equation}
for all $(m,n)$.

If we gauge $\nabla^2$ and $(\nabla^2+R^2 + 2\mE +M)$ with the generalized Gaussian, we obtain
\begin{equation}
\label{berekexpdelta}
\exp(R^2/2) \nabla^2\exp{(-R^2/2)}= \nabla^2+R^2 -2\mE -M
\end{equation}
and
\begin{equation}
\label{CHexp}
\exp(R^2/2)(\nabla^2+R^2 + 2\mE +M) \exp{(-R^2/2)}=\nabla^2.
\end{equation}

In the full superspace, lemma \ref{Laplfischer} takes the following form.

\begin{lemma}
Let $H_k \in \cH_k$ and $M \not \in -2 \mN$. Then for all $i,j,k \in \mN$ one has
\[
\nabla^{2i}(R^{2j} H_k) = \left\{
\begin{array}{l}
c_{i,j,k} R^{2j-2i} H_k, \quad i \leq j\\
\vspace{-2mm}\\
0, \quad i > j
\end{array}
\right.
\]
with
\[
c_{i,j,k} = 4^{i} \frac{j!}{(j-i)!} \frac{\Gamma(k+M/2+j)}{\Gamma(k+M/2+j-i)}.
\]
\label{laplonpieces}
\end{lemma}

The integration we use on a full superspace is defined by
\begin{equation}
\label{superint}
\int_{\mR^{m | 2n}} = \int_{\mR^m} d V(\ux) \int_B=\int_B \int_{\mR^m} d V(\ux).
\end{equation}
The Schwartz space in superspace is given by $\cS(\mR^m)_{m|2n}=\cS(\mR^m)\otimes\Lambda_{2n}$. In \cite{DBS9}, a Fourier transform in superspace was introduced for $f\in\cS(\mR^m)_{m|2n}$ by
\[
\cF_{m | 2n}^{\pm}(f) = (2 \pi)^{-\frac{M}{2}} \int_{\mR^{m | 2n},x} \exp{(\pm i \langle \bold{x},\bold{y} \rangle)}f(\bold{x})
\]
with
\begin{equation}
\label{innerprodvectorssuper}
\langle \bold{x} , \bold{y} \rangle = \sum_{i=1}^{m} x_i y_i -\frac{1}{2} \sum_{j=1}^{n}({x \grave{}}_{2j-1}{y \grave{}}_{2j} - {x \grave{}}_{2j} {y \grave{}}_{2j-1})= \langle\ux,\uy\rangle+ \langle \uxb,\uyb\rangle.
\end{equation}
In this notation, ${x \grave{}}_{i}$ and ${y \grave{}}_{j}$ generate a Grassmann algebra $\Lambda_{4n}$. In particular, ${x \grave{}}_{i}$ and ${y \grave{}}_{j}$ are mutually anti-commutative and $\langle \bold{x},\bold{y} \rangle$ is symmetric.

Note that this Fourier transform can also be defined as the operator exponential
\begin{equation}
\label{FTSuperExp}
\cF_{m | 2n}^{\pm} =e^{\mp \frac{ i \pi M}{4} } e^{\pm \frac{i \pi}{4}(-\nabla^2 +R^2)}.
\end{equation}
The super Fourier transform satisfies similar properties as the classical bosonic Fourier transform defined in (\ref{classFT}). The inverse of $\cF^+$ is given by $\cF^-$, i.e.
\[
\cF_{m | 2n}^{\pm} \circ \cF_{m | 2n}^{\mp} = \mbox{id}_{\, \cS(\mR^m)_{m|2n}}.
\]
The Fourier transform of the spherical Hermite functions is
\[
\cF_{m | 2n}^{\pm} (\varphi_{j,k,l}(\bold{x}) )=(\pm i )^{2j+ k} \varphi_{j,k,l}(\bold{y}).
\]

The extension of the Fourier transform from $\cS(\mR^m)_{m|2n}$ to $L_{2}(\mR^m)_{m|2n} =L_2(\mR^m)\otimes\Lambda_{2n}$ is trivial because clearly $\cF_{m|2n}^\pm=\cF_{m|0}^\pm\circ\cF_{0|2n}^\pm$.

When $M\not\in-2\mN$ or when $m=0$, we define the fractional Fourier transform $\cF^{\alpha}_{m|2n}$ on $\cS(\mR^m)\otimes\Lambda_{2n}$,  by its action on the basis functions (see \cite{DBS9}): 
\begin{equation}
\label{fracFourbasis}
\cF^{\alpha}_{m|2n} (\varphi_{j,k,l}(\bold{x})) = e^{  i \alpha (2j+k)} \varphi_{j,k,l}(\bold{y}),
\end{equation}
where $\alpha \in [-\pi/2,\pi/2]$. The fractional Fourier transform thus rotates the basis functions over a multiple of the angle $\alpha$. In the limit case $\alpha = \pm \pi/2$, the fractional Fourier transform reduces to the ordinary Fourier transform. We have the following integral representation (see \cite{DBS9}, theorem 12).

\begin{theorem}
On $\cS(\mR^m)_{m|2n}$, the fractional Fourier transform is given by
\[
\cF^{\alpha}_{m|2n}(f(\bold{x}))= c \int_{\mR^{m|2n},x}   \exp{\frac{ 4 e^{i \alpha} \langle \bold{x},\bold{y} \rangle - (1+ e^{2i \alpha})(R^2 + R_{\bold{y}}^2)}{2- 2e^{2i \alpha}}} f(\bold{x}),
\]
with $c= \left(\pi (1- e^{2i \alpha})\right)^{-M/2}$.
\label{fracfourthm}
\end{theorem}

\begin{remark}
Note that the sign convention we use in the definition of the Fourier transform (see formula (\ref{innerprodvectorssuper})) is slightly different than the one used in \cite{DBS9}.
\end{remark}

Finally, we repeat some important facts about spherical harmonics in superspace. For proofs we refer the reader to \cite{DBE1}.
\begin{lemma}
\label{polythm}
If $0\le q \le n$ and $0\le k\le n-q$, there exists a homogeneous polynomial $f_{k,p,q}=f_{k,p,q}(r^2,\theta^2)$ (unique up to a multiplicative constant) of total degree $k$ such that $f_{k,p,q} \cH_p^b \otimes \cH_q^f \neq 0$ and $\nabla^2 (f_{k,p,q} \cH_p^b \otimes \cH_q^f) = 0$.
This polynomial is given explicitly by
\[
 f_{k,p,q}=\sum_{s=0}^ka_sr^{2k-2s}\theta^{2s},\quad a_s=\binom{k}{s}\frac{(n-q-s)!}{\Gamma (\frac{m}{2}+p+k-s)}\frac{\Gamma(\frac{m}{2}+p+k)}{(n-q-k)!}.
\]
\end{lemma}

In particular $f_{0,p,q}=1$ holds. To simplify the subsequent formulas, we have used a different normalization of the $f_{k,p,q}$ as in \cite{DBE1}. Using these polynomials we can obtain a full decomposition of the space of spherical harmonics of degree $k$.

\begin{theorem}[Decomposition of $\cH_k$]
\label{decompintoirreps}
Under the action of $SO(m) \times Sp(2n)$ the space $\cH_k$ decomposes as
\[
\cH_{k} = \bigoplus_{j=0}^{\min(n, k)} \bigoplus_{l=0}^{\min(n-j,\lfloor \frac{k-j}{2} \rfloor)} f_{l,k-2l-j,j} \cH^b_{k-2l-j} \otimes \cH^f_{j},
\]with $f_{l,k-2l-j,j}$ the polynomials determined in lemma \ref{polythm}.
\end{theorem}

The integration on superspace (\ref{superint}) for $m\not=0$ is linked with an integration over the supersphere defined by the following Pizzetti formula (see \cite{DBS5,DBE1}):
\begin{equation}
\int_{SS} f =  \sum_{k=0}^{\infty}  \frac{2 \pi^{M/2}}{2^{2k} k!\Gamma(k+M/2)} (\nabla^{2k} f)(0), \quad f \in \cP.
\label{defintss}
\end{equation}
The supersphere is formally defined as the algebraic object $R^2-1=0$. It can be proven that (\ref{defintss}) is, up to a constant, the unique linear functional on $\cP$ satisfying $\int_{SS}R^2 f=\int_{SS}f$, which is invariant under the group $SO(m)\times Sp(2n)$ and which makes spherical harmonics of different degree `orthogonal' (see \cite{DBE1}, \cite{CDBS1}).  In fact, this orthogonality condition can be made even stronger (see \cite{DBE1}, theorem 8):

\begin{theorem}
\label{integorth}
One has that $f_{i,p,q} \cH^b_{p} \otimes \cH^f_{q} \quad  \bot \quad f_{j,r,s} \cH^b_{r} \otimes \cH^f_{s}$, meaning 
\[
\int_{SS} \left(f_{i,p,q} \cH^b_{p} \otimes \cH^f_{q}\right)  \left(f_{j,r,s} \cH^b_{r} \otimes \cH^f_{s}\right) =0
\]
with respect to the Pizzetti integral, if and only if $(i,p,q) \neq (j,r,s)$.
\end{theorem}

In case $M\not\in-2\mN$, for $P_k$ a homogeneous polynomial of degree $k$ we also have
\begin{equation}
\label{superint2}
\int_{\mR^{m|2n}}P_k\exp (-R^2) =\frac{1}{2}\Gamma (\frac{k+M}{2})\int_{SS} P_k.
\end{equation}
Finally, the supersphere integration given by (\ref{defintss}) can be extended to more general (non-polynomial) functions as follows (see \cite{CDBS1}, theorem 8) 
\begin{eqnarray}
\label{SSintnieuw}
\int_{SS}f&=& \sum_{j=0}^n\int_{\mS^{m-1}} d \sigma \int_B\frac{(-1)^j\theta^{2j}}{j!}\left[(\frac{\partial}{\partial r^2})^jr^{m-2}f\right]_{r=1}.
\end{eqnarray}

\subsection{Schr\"odinger equations in superspace}
\label{SEsuper}

Schr\"odinger equations in superspace are equations of the type
\begin{equation}
\label{SchrodEq}
-\frac{\nabla^2}{2} \psi + V \psi = E \psi
\end{equation}
with wave function $\psi \in L_{2}(\mR^{m})_{m|2n}$ and the energy $E$ a complex number. For more background on Schr\"odinger equations in superspace and supersymmetric quantum mechanics, see e.g. \cite{MR0780110, MR0723957, MR0952024, MR2277080, MR2025382, Witten, MR0683171}. The potential $V$ is expressed in terms of commuting and anti-commuting variables. In the case $m=0$ this is a purely algebraic eigenvalue problem. When $m \neq 0$, equation (\ref{SchrodEq}) is equivalent with a system of PDEs, as can be observed by expanding $\psi$ in the anti-commuting variables.

Several authors have studied explicit examples of such Schr\"odinger equations. The (purely fermionic) harmonic oscillator was studied in \cite{MR830398}. Anharmonic extensions were studied in \cite{MR967935,MR1019514,MR1032208}. In \cite{ZHANG} the hydrogen atom in superspace (or quantum Kepler problem) was studied using Lie superalgebra techniques. Also the delta potential has been studied, see \cite{DBS8}. 

In this paper we consider general potentials of the form $V(R^{2})$, with $V$ a polynomial. The simplest case is then the harmonic oscillator described by the hamiltonian
\begin{equation}
\label{hamiltoniaan}
H= \frac{1}{2}(-\nabla^2+R^2) = \sum_{i=1}^m a_i^+ a_i^- +\sum_{i=1}^{2n} b_i^+ b_i^- + \frac{M}{2}
\end{equation}
with
\[
\begin{array}{llll}
a^+_i = \frac{\sqrt{2}}{2}(x_i - \pI ) &\quad&a^-_i = \frac{\sqrt{2}}{2}(x_i+\pI)\\
\vspace{-1mm}\\
b^+_{2i} = \frac{1}{2}({x \grave{}}_{2i} + 2 \partial_{{x \grave{}}_{2i-1}})&\quad&b^-_{2i} =\frac{1}{2} ({x \grave{}}_{2i-1} + 2 \partial_{{x \grave{}}_{2i}})\\
\vspace{-1mm}\\
b^+_{2i-1} = \frac{1}{2}({x \grave{}}_{2i-1} - 2 \partial_{{x \grave{}}_{2i}})&\quad&b^-_{2i-1} = \frac{1}{2}(-{x \grave{}}_{2i} + 2 \partial_{{x \grave{}}_{2i-1}})\\
\vspace{-1mm}\\
\end{array}
\]
the bosonic and fermionic creation and annihilation operators. The spherical Hermite functions defined in (\ref{CliffordHermiteFunctionssuper}) form a basis of eigenvectors of $H$ for $\cP\exp(-R^2/2)$, satisfying (see \cite{DBS3})
\begin{equation}
H \varphi_{j,k,l} = \left( \frac{M}{2} + (2j+k)\right) \varphi_{j,k,l}.
\label{HOCH} 
\end{equation}

\section{Hermite polynomials in Grassmann algebras}
\setcounter{equation}{0}
\label{FermionicMehler}

\subsection{Inner product on a Grassmann algebra}

We first define two vector space isomorphisms of the Grassmann algebra $\Lambda_{2n}$.
\begin{definition}
\label{defmaptilde}
The transformation $\widetilde{.}:\Lambda_{2n}\to \Lambda_{2n} $ is a linear transformation defined by
\begin{eqnarray*}
\widetilde{{x \grave{}}_{2i-1}}&=&{x \grave{}}_{2i}\\
\widetilde{{x \grave{}}_{2i}}&=&-{x \grave{}}_{2i-1}\\
\widetilde{ab}&=&\widetilde{b}\widetilde{a}, \qquad a,b \in \Lambda_{2n}.
\end{eqnarray*}
\end{definition}

This transformation satisfies the following property.
\begin{lemma}
\label{Sptilde}
For each $B\in Sp(2n)$, define $D\in Sp(2n)$ as $D=J^TBJ$ (as matrix multiplications). The transformation $\widetilde{\cdot}$ intertwines with the action of $Sp(2n)$ in the following way:
\[
\widetilde{\cdot}\circ B= D\circ\widetilde{\cdot}.
\]
\end{lemma}
\begin{proof}
For a monomial ${x\grave{}}_A={x\grave{}}_{i_1}\cdots {x\grave{}}_{i_k}$, definition \ref{defmaptilde} yields
\begin{eqnarray*}
\widetilde{{x\grave{}}_A}=\sum_{j_1,\cdots,j_k}J_{i_1j_1}\cdots J_{i_kj_k}{x\grave{}}_{j_k}\cdots {x\grave{}}_{j_1}.
\end{eqnarray*}
The definition of the action of $Sp(2n)$ in formula \eqref{actieOSp} implies
\begin{eqnarray*}
B{x\grave{}}_A&=&\sum_{j_1,\cdots,j_k}B_{j_1i_1}\cdots B_{j_ki_k}{x\grave{}}_{j_1}\cdots {x\grave{}}_{j_k}.
\end{eqnarray*}
Combining these transformations yields
\begin{eqnarray*}
\widetilde{B{x\grave{}}_A}&=&\sum_{j_1,\cdots,j_k}B_{j_1i_1}\cdots B_{j_ki_k}\sum_{l_1,\cdots, l_k}J_{j_1l_1}\cdots J_{j_k l_k}{x\grave{}}_{l_k}\cdots {x\grave{}}_{l_1}
\end{eqnarray*}
and
\begin{eqnarray*}
D\widetilde{{x\grave{}}_A}&=&\sum_{j_1,\cdots,j_k}J_{i_1j_1}\cdots J_{i_kj_k}\sum_{l_1,\cdots,l_k}D_{l_1j_1}\cdots D_{l_kj_k}{x\grave{}}_{l_k}\cdots {x\grave{}}_{l_1}.
\end{eqnarray*}
For the last two expressions to be identical, $DJ^T=J^TB$ needs to hold, which is equivalent with $D=J^TBJ$.
\end{proof}

\begin{definition}
\label{defstar}
The star map $\ast$ maps monomials ${x \grave{}}_A = {x \grave{}}_1^{\, \alpha_1} \ldots {x \grave{}}_{2n}^{\, \alpha_{2n}}$ of degree $k$ to monomials $\ast {x \grave{}}_A = \pm 2^{k-n} {x \grave{}}_1^{1-\alpha_1} \ldots {x \grave{}}_{2n}^{1-\alpha_{2n}}$ of degree $(2n-k)$ where the sign is chosen such that
\[
{x \grave{}}_A (\ast {x \grave{}}_A )= 2^{k-n}  {x \grave{}}_{1} \ldots {x \grave{}}_{2n}.
\]
By linearity, $\ast$ is extended to the whole of $\Lambda_{2n}$.
\end{definition}
It is easy to check that $\ast \ast {x \grave{}}_A = (-1)^{k (k-2n)} {x \grave{}}_A$, so $\ast$ behaves very similarly as the Hodge star map acting on the space of differential forms on a Riemannian manifold. Because the dimension $(-2n)$ in our case is always even, we obtain 
\[\ast\ast {x\grave{}}_A=(-1)^k{x\grave{}}_A=\widetilde{\widetilde{{x\grave{}}_A}}.
\]
Note that $\ast$ and $\widetilde{.}$ are vector space isomorphisms on $\Lambda_{2n}$. It is also easy to check that $\widetilde{.}$ leaves $\cH_{k}^{f}$ invariant.

Using the star map we define the following inner product on $\Lambda_{2n}$ (where we use the bra-ket notation for convenience):

\begin{definition}
\label{inprodGrass}
The inner product $\langle.|.\rangle:\Lambda_{2n}\times \Lambda_{2n}\to \mC$ is given by
\begin{eqnarray*}
\langle f|g \rangle &=& \int_{B,x} f (\ast \overline{g})\\
&=& \pi^{-n} \partial_{{x \grave{}}_{2n}} \ldots \partial_{{x \grave{}}_{1}} f (\ast \overline{g})\\
&=& \frac{1}{(2\pi)^n} \sum_A 2^{|A|} f_A \overline{g_{A}},
\end{eqnarray*}
with $\overline{\, \cdot\, }$ the standard complex conjugation, $f=\sum_{A}f_A{x\grave{}}_A$ and  $g=\sum_{A}g_A{x\grave{}}_A$. 
\end{definition}

This inner product, in a different formulation, was also used in \cite{MR2025382}. When we will introduce new inner products in superspace (see section \ref{supersuper}), we will use the notation $\langle.|.\rangle_{\Lambda_{2n}}$ for the inner product in definition \ref{inprodGrass} to avoid confusion. The fact that $\langle.|.\rangle$ is an inner product follows immediately from the last expression in the definition. This immediately implies the following lemma.

\begin{lemma}
\label{triviaalstelemma}
If for $p\in \Lambda_{2n}$ it holds that $\int_B pq=0$ for every $q$ $\in \Lambda_{2n}$, then $p=0$.
\end{lemma}
\begin{proof}
Using the fact that the star map is an isomorphism, $\int_B pq=0$ for every $q$ $\in \Lambda_{2n}$ implies 
\[
\int_B p (\ast \overline{p})=0= \langle p | p\rangle.
\]
This is only possible when $p=0$.
 \end{proof}

It is easy to see that any inner product on the Grassman algebra $\Lambda_{2n}$ can be written similarly to definition \ref{inprodGrass} as $\langle f|g \rangle=\sum_{A,B} f_A P_{AB}\overline{ g_B}$ with $P$ a hermitian positive definite $2^{2n}\times2^{2n}$ matrix. This is an inner product of the form  $\int_{B,x} f (\Phi\overline{g})$, with $\Phi$ a different isomorphism on the Grassman algebra.

The inner product in definition \ref{inprodGrass} is not covariant as the classical bosonic inner product $\int_{\mR^m} f\overline{g} dV(\ux)$, because the star map does not behave covariantly for transformations with determinant equal to one. However, it can be shown that there are no covariant positive definite inner products on $\Lambda_{2n}$. Before doing so, we reobtain the well-known fact that the Berezin integral itself does behave covariantly.

\begin{lemma}
\label{Berezincov}
Let $A$ be a transformation on the generators of the Grassmann algebra given by ${x\grave{}}_k=\sum_{j=1}^{2n}A_{kj}{y\grave{}}_j$. Then one has
\begin{equation*}
\int_{B,y} f(\uxb (\uyb)) \equiv \int_{B,y} f(A.\uyb) =\det (A) \int_{B,x}f(\uxb).
\end{equation*}
\end{lemma}

\begin{proof}
The transformation preserves the degree of $f$, so we only have to consider the case $f(\uxb)={x \grave{}}_{1} \ldots {x \grave{}}_{2n}$. Because of the anticommutation rules we immediately have $f(\uxb(\uyb))=\det (A) {y \grave{}}_{1} \ldots {y \grave{}}_{2n}$. As $\int_{B,y}$ is defined by $ \pi^{-n} \partial_{{y \grave{}}_{2n}} \ldots \partial_{{y \grave{}}_{1}}$, the lemma follows.
 \end{proof}

So for transformations with determinant one the integration is independent of the choice of coordinates.

\begin{remark}
From the calculation above it is clear that the Jacobian determinant appears on the other side in comparison to the bosonic case. Therefore, the above formula for substitution in the Berezin integral is usually written after multiplying  both sides with $1/\det (A)$. The factor $1/\det (A)$ is then called the Berezinian (see \cite{MR732126}), the fermionic equivalent of the Jacobian. In general the Berezinian is a superdeterminant, but in the purely fermionic case this is equivalent to the inverse.
\end{remark}

\begin{proposition}
\label{geencov}
There are no inner products on the Grassmann algebra $\Lambda_{2n}$ with the property that
\begin{eqnarray*}
\langle f(\uxb) |g(\uxb)\rangle_x &=& \langle f(\uxb(\uyb) |g(\uxb(\uyb))\rangle_y
\end{eqnarray*}
for transformations $A$ as in lemma \ref{Berezincov} with $\det (A)=1$.
\end{proposition}

\begin{proof}
We already mentioned that every inner product on the Grassmann algebra is of the form
\[
\langle f|g\rangle=\sum_{A,B}f_AP_{AB} \overline{g_B},
\]
for some hermitian positive definite $2^{2n}\times 2^{2n}$ matrix $P$. Now we only consider the case with $f$ and $g$ homogeneous of degree one. Then we get a submatrix of $P$, the hermitian positive definite $2n\times 2n$ matrix $Q$. For the inner product to be covariant, $Q$ has to satisfy $R^H QR=Q$ for every matrix $R$ with determinant one and with $\cdot^H$ the hermitian conjugate. Basic linear algebra shows that such a non-zero $Q$ does not exist. 
 \end{proof}

Proposition \ref{geencov} can be made even stronger.  By similar arguments one can show that there are no inner products on the Grassmann algebra which are invariant under symplectic transformations. In the subsequent proposition \ref{Spinprod} the behavior of the inner product with respect to the symplectic group will be studied.

Now we derive the properties of the inner product given in definition \ref{inprodGrass}. We start with the following lemma, which shows that the adjoint of ${x\grave{}}_j$ is given by the usual adjoint in supersymmetric quantum mechanics.
\begin{lemma}
\label{adjoints}
The adjoint of $\partial_{{x \grave{}}_{j}}$ with respect to the inner product $\langle . | . \rangle $ on $\Lambda_{2n}$ is given by ${x \grave{}}_{j}/2$. This property determines the inner product uniquely, up to a multiplicative constant.
\end{lemma}
\begin{proof}
We need to prove that $\langle \partial_{{x \grave{}}_{j}}f|g \rangle = \langle f| \frac{{x \grave{}}_{j}}{2} g \rangle$.
Due to linearity, it suffices to take $f = a {x \grave{}}_j {x \grave{}}_A$, $g = b {x \grave{}}_A$ with $a,b \in \mC$. Then
\[
\langle  \partial_{{x \grave{}}_{j}}f|g \rangle = \int_{B,x} a {x \grave{}}_A (\ast \overline{b} {x \grave{}}_A)= \frac{\pi^{-n}}{2^n}  2^{|A|} a \overline{b} 
\]
and on the other hand
\[
\langle  f| {x \grave{}}_{j}g \rangle = \int_{B,x} a {x \grave{}}_{j}{x \grave{}}_A (\ast \overline{b} {x \grave{}}_{j}{x \grave{}}_A)
= \frac{\pi^{-n}}{2^n}  2^{|A| +1} a \overline{b} ,
\]
proving the first part of the lemma. Now suppose we have a different inner product on $\Lambda_{2n}$, $(.|.)$, for which the same property holds. For two monomials ${x \grave{}}_A = {x \grave{}}_{1}^{\, \alpha_1} \ldots {x \grave{}}_{2n}^{\, \alpha_{2n}}$ of degree $k$ and ${x \grave{}}_B = {x \grave{}}_1^{\, \beta_1} \ldots {x \grave{}}_{2n}^{\, \beta_{2n}}$ of degree $l$ (with $k\ge l$) we find
\begin{eqnarray*}
({x \grave{}}_A|{x \grave{}}_B)&=&2^k(1|\partial_{{x \grave{}}_{2n}}^{\, \alpha_{2n}} \ldots \partial_{{x \grave{}}_{1}}^{\, \alpha_{1}}{x \grave{}}_B).
\end{eqnarray*}
Now, $\partial_{{x \grave{}}_{2n}}^{\, \alpha_{2n}} \ldots \partial_{{x \grave{}}_{1}}^{\, \alpha_{1}}{x \grave{}}_B$ is zero when $k>l$, or when $k=l$ and ${x \grave{}}_A\not={x \grave{}}_B$. This means that we find $({x \grave{}}_A,{x \grave{}}_B)=0$ when $A\not= B$ and $({x \grave{}}_A,{x \grave{}}_A)=2^k(1|1)$, which corresponds to definition \ref{inprodGrass}.
 \end{proof}

Contrary to the bosonic $L_2$ inner product, $\partial_{{x\grave{}}_j}$ is defined on the entire space $\Lambda_{2n}$, which simplifies the notion of an adjoint operator. For the sequel, we need the adjoints of the generators of the $\mathfrak{sl}_{2}$ algebra.

\begin{corollary}
\label{hermtoeg}
With respect to the inner product $\langle . | . \rangle $ on $\Lambda_{2n}$ the adjoints of $\theta^2$, $\nabla^2_f$ and $\mE_f-n$ are given by
\begin{equation*}
(\theta^2)^\dagger=-\nabla^2_f, \qquad (\nabla^2_f)^\dagger=-\theta^2, \qquad (\mE_f-n)^\dagger=(\mE_f-n).
\end{equation*}
\end{corollary}

\begin{proof}
The first two adjoints follow immediately from the previous lemma, e.g. $({x \grave{}}_{2j-1}{x \grave{}}_{2j})^\dagger={x \grave{}}_{2j}^\dagger {x \grave{}}_{2j-1}^\dagger=-4\partial_{{x \grave{}}_{2j-1}}\partial_{{x \grave{}}_{2j}}$.
The last one follows from the first two and formula (\ref{fermEuler}):
\[
(\mE_f-n)^\dagger=\frac{1}{4}[\nabla^2_f,\theta^2]^\dagger=\frac{1}{4}[(\theta^2)^\dagger,(\nabla^2_f)^\dagger]=\frac{1}{4}[-\nabla^2_f,-\theta^2].
\]
\end{proof}

Recall that the hamiltonian of the fermionic harmonic oscillator is given by (see (\ref{hamiltoniaan}))
\[
H = (-\nabla^2_f +\theta^2)/2.
\]
Using corollary \ref{hermtoeg} we immediately obtain that $H$ is self-adjoint, hence a notation such as $\langle f|H |g \rangle$ makes sense. 

It would of course be preferable to have an inner product for which $(\theta^2)^\dagger=\theta^2$ and $(\nabla^2_f)^\dagger=\nabla^2_f$, similar to the bosonic $L_2$ inner product. With such an inner product other hamiltonians would be hermitian too. An interesting class of relevant hamiltonians, as already indicated in section \ref{SEsuper}, is of the form $\Delta_f/2 +V(\theta^2)$, with $V$ a polynomial. Several examples have already been studied in \cite{MR967935,MR1019514,MR1032208}. It is however easy to see that such an inner product does not exist.

\begin{proposition}
\label{onmogelijkinprod}
There are no inner products on the Grassman algebra $\Lambda_{2n}$ for which $(\theta^2)^\dagger=\theta^2$.
\end{proposition}

\begin{proof}
If multiplication with $\theta^2$ is a hermitian operation then 
\begin{eqnarray*}
\langle \theta^{2n}|\theta^{2n}\rangle &=& \langle \theta^{2n-2}|\theta^{2n+2} \rangle=0
\end{eqnarray*}
as $\theta^{2n+2}=0$. Hence $\langle.|.\rangle $ is not positive definite and thus not an inner product. 
 \end{proof}

This proposition does not really form a limitation for negative dimensional quantum mechanics per se, as can be seen from the isotropic anharmonic oscillator. In \cite{MR967935} it was calculated that the eigenvalues for the hamiltonian $H=-\nabla^2_f+\theta^2-\lambda \theta^4$ (with $\lambda$ real) can be complex. Such a hamiltonian can therefore never be hermitian with respect to an inner product.

\vspace{3mm}
We now investigate the action of the star map in more detail. Therefore we first state some useful properties of the Berezin integral.
\begin{lemma}
\label{Berezin}
If $h\in \Lambda_{2n}^{k}$ and $f,g \in \Lambda_{2n}$, then the following relations hold
\begin{align*}
&(i) \int_B (\partial_{{x \grave{}}_{j}} f )h=(-1)^k\int_B f\partial_{{x \grave{}}_{j}} h &\quad&(iii) \int_B (\theta^2 f) g=\int_B f\theta^2  g\\
&(ii) \int_B (\nabla^2_f f) g=\int_B f\nabla^2_f  g &\quad&(iv) \int_B ((\mE_f-n)f)g=-\int_B f(\mE_f-n)g.
\end{align*}
\end{lemma}

\begin{proof}
The first property follows from $\int_B \partial_{{x \grave{}}_{j}}=0$ and the fact that we only have to consider $f$ of degree $2n-k+1$. $(ii)$ immediately follows from $(i)$. $(iii)$ is trivial and $(iv)$ follows from $(ii)$ and $(iii)$ by equation (\ref{fermEuler}).
 \end{proof}

Using the previous lemma we obtain the following calculation rules for the star map.

\begin{lemma} 
\label{eigstar} 
If $f\in \Lambda_{2n}^{k}$ and $g\in \Lambda_{2n}$, then the following relations hold
\begin{align*}
& (i)*{x \grave{}}_{j}f=(-1)^k2\partial_{{x \grave{}}_{j}}*f&\quad &(iv)*\theta^2 g = -\nabla^2_f *g\\
& (ii)*\partial_{{x \grave{}}_{j}}f=-(-1)^k\frac{1}{2}{x \grave{}}_{j}*f&\quad &(v)*(\mE_f-n)g= -(\mE_f-n)* g\\
& (iii)*\nabla^2_f g= -\theta^2 * g&\quad &(vi)*1=\frac{1}{2^n}{x \grave{}}_{1}\ldots{x \grave{}}_{2n}=\frac{(-1)^n\theta^{2n}}{2^n n!}.
\end{align*}
\end{lemma}

\begin{proof}
The proof follows easily from the lemmas \ref{triviaalstelemma}, \ref{adjoints} and \ref{Berezin}. As an example we prove $(i)$. For every $g\in\Lambda_{2n}$, we have
\begin{eqnarray*}
\int_B g(\ast {x\grave{}}_jf)&=&2\int_B(\partial_{{x\grave{}}_j}g)(\ast f)\\
&=&2(-1)^{2n-k}\int_Bg(\partial_{{x\grave{}}_j}\ast f).
\end{eqnarray*}
This implies $(i)$ by lemma \ref{triviaalstelemma}. 
 \end{proof}

Now we will calculate the action of the star map on the spherical Hermite functions defined in (\ref{CliffordHermiteFunctions}). We start with two auxiliary results.

\begin{lemma}
\label{astHk}
For $H_k \in \cH_k^f$, the following holds:\\
\begin{equation*}
*H_k=\widetilde{H}_k \frac{(-1)^{n-k}\theta^{2n-2k}}{2^{n-k}(n-k)!}.
\end{equation*}
\end{lemma}

\begin{proof}
We prove this lemma by induction on $n$. For $n=1$ the result is trivial. If it holds for $n-1$, then theorem \ref{basisrecursief} provides us with a useful basis for $\cH_k^f$ in $\Lambda_{2n}$. We take $\Lambda_{2n-2}$ the Grassmann algebra without ${x\grave{}}_1$ and ${x\grave{}}_2$ and put $\theta^{2}_{2n-2} =- \sum_{j=2}^n {x\grave{}}_{2j-1} {x\grave{}}_{2j}$. Using definition \ref{defstar} we find that for a monomial ${x\grave{}}_A\in\Lambda_{2n-2}$ of degree $k$, $\ast_{n}{x\grave{}}_A=(\ast_{n-1}{x\grave{}}_A)\frac{{x\grave{}}_1{x\grave{}}_2}{2}$. By linearity this holds for every element of $\Lambda_{2n-2}$. In theorem \ref{basisrecursief} there are 3 different types of spherical harmonics. For the first type (namely $H_{k} \in \cH_k^f({x\grave{}}_3, \ldots, {x\grave{}}_{2n})$) we find, using the induction hypothesis,
\begin{eqnarray*}
\ast_nH_k&=&(\ast_{n-1}H_k)\frac{{x\grave{}}_1{x\grave{}}_2}{2}\\
&=&\widetilde{H}_k \frac{(-1)^{n-k-1}\theta^{2n-2k-2}_{2n-2}}{2^{n-k-1}(n-k-1)!}\frac{{x\grave{}}_1{x\grave{}}_2}{2}\\
&=&\widetilde{H}_k \frac{(-1)^{n-k}\theta^{2n-2k}}{2^{n-k}(n-k)!}.
\end{eqnarray*}
Indeed, we have
\[
(-{x\grave{}}_1{x\grave{}}_2 + \theta_{2n-2}^{2})^{n-k} = \theta_{2n-2}^{2n-2k} - (n-k) {x\grave{}}_1{x\grave{}}_2 \, \theta_{2n-2}^{2n-2k-2}
\]
and $\widetilde{H}_k\theta^{2n-2k}_{2n-2} = 0$ because of equation (\ref{Fischer}).

For the two other types the lemma follows from the result for the first type and the calculation rules in lemma \ref{eigstar}.
 \end{proof}

\begin{lemma}
\label{expstar}
For all $H_k \in \cH_k^f$ and for $i+k\le n$ one has
\begin{eqnarray*}
& &(i)*\theta^{2i}H_k= (-1)^{n-k}2^i i! \frac{\theta^{2n-2k-2i}}{2^{n-k-i}(n-k-i)!} \widetilde{H}_k \\
& &(ii)*H_k\exp(-\theta^2/2)=\widetilde{H}_k\exp(-\theta^2/2).
\end{eqnarray*}
\end{lemma}

\begin{proof}
Using lemma \ref{Laplfischer}, lemma \ref{eigstar} and lemma \ref{astHk} we obtain
\begin{eqnarray*}
*\theta^{2i}H_k&=&(-1)^{i+n-k}\nabla_f^{2i} \frac{\theta^{2n-2k}}{2^{n-k}(n-k)!}\widetilde{H}_k\\
&=&(-1)^{n-k}4^i\frac{(n-k)!}{(n-k-i)!}\frac{(n+i-k-n+k)!}{(n-k-n+k)!}\frac{\theta^{2n-2k-2i}}{2^{n-k}(n-k)!}\widetilde{H}_k\\
&=&(-1)^{n-k}4^i\frac{i!}{(n-k-i)!}\frac{\theta^{2n-2k-2i}}{2^{n-k}}\widetilde{H}_k.
\end{eqnarray*}
This proves the first formula. Using this result we then have
\begin{eqnarray*}
*H_k \exp(-\theta^2/2)&=&\sum_{i=0}^{n-k} \frac{1}{2^ii!}*(-1)^i\theta^{2i} H_k\\
&=&\sum_{i=0}^{n-k}\frac{(-1)^{n-k-i}\theta^{2n-2k-2i}}{2^{n-k-i}(n-k-i)!} \widetilde{H}_k
=\sum_{j=0}^{n-k}\frac{(-1)^j\theta^{2j}}{2^jj!}\widetilde{H}_k.
\end{eqnarray*}
 \end{proof}

We can now explicitly state the action of the star map on the spherical Hermite functions defined in (\ref{CliffordHermiteFunctions}).

\begin{theorem}
\label{CHstar}
If $H_k^{(l)}\in \cH_k^f$ and $j+k\le n$, then the following holds for the $\varphi^{f}_{j,k,l}$ defined in equation (\ref{CliffordHermiteFunctions}):
\begin{eqnarray*}
*L_j^{-n+k-1}(\theta^2) H_k^{(l)}\exp(-\theta^2/2)=(-1)^j L_j^{-n+k-1}(\theta^2)\widetilde{H}_k^{(l)}\exp(-\theta^2/2).
\end{eqnarray*}

\end{theorem}
\begin{proof}
This follows from the lemmas \ref{eigstar} and \ref{expstar} $(ii)$ as follows
\begin{eqnarray*}
*\varphi^{f}_{j,k,l} &=& *(-\nabla^2_{f} -\theta^2 + 2\mE -2n)^j H_k^{(l)} \exp{(-\theta^2/2)} \\
&=& (\theta^2+ \nabla^2_{f} - 2\mE +2n)^j *H_k^{(l)} \exp{(-\theta^2/2)}\\
&=&(-1)^j(-\nabla^2_f-\theta^2+2\mE-2n)^j\widetilde{H}_k^{(l)}\exp(-\theta^2/2)\\
&=&(-1)^j 2^{2j}j! L_j^{-n+k-1}(\theta^2)\widetilde{H}_k^{(l)}\exp(-\theta^2/2).
\end{eqnarray*}
 \end{proof}

Although the inner product is not invariant under the symplectic group, it does behave canonically with respect to it.
\begin{proposition}
\label{Spinprod}
For $f,g\in\Lambda_{2n}$ and $A\in Sp(2n)$, with action as defined in formula (\ref{actieOSp}), the following relation holds
\begin{eqnarray*}
\langle Af|g\rangle &=&\langle f|A^Tg\rangle,
\end{eqnarray*}
which implies $A^\dagger=A^T$, with $A^T$ the matrix transpose of $A$. This is equivalent with
\begin{eqnarray*}
\langle AJf|JAg\rangle&=&\langle f|g\rangle,
\end{eqnarray*}
for all $A\in Sp(2n)$.
\end{proposition}

\begin{proof}
Lemma \ref{Sptilde} and \ref{expstar} imply that $\ast A g= (J^TAJ)\ast g$ for $A\in Sp(2n)$. This implies that
\begin{eqnarray*}
\langle (AJ)f|(JA)g\rangle&=&\int_B (AJf)(\ast (JA\overline{g}))\\
&=&\int_B (AJf)(AJ\ast \overline{g})=\int_B AJf(\ast \overline{g}).
\end{eqnarray*}
Lemma \ref{Berezincov} then yields
\begin{eqnarray*}
\langle (AJ)f|(JA)g\rangle={\det(AJ)^T}\int_B f(\ast \overline{g})=\int_B f(\ast \overline{g}),
\end{eqnarray*}
since $\det A=1$ for all $A\in Sp(2n)$. It is easily checked that the map $A\to AJ$ is a bijection of $Sp(2n)$. Since for $A\in Sp(2n)$, the relation $(AJ)^{-1}=(JA)^{T}$ holds, the claim $\langle AJf|JAg\rangle=\langle f|g\rangle$ for all $A\in Sp(2n)$, is equivalent with stating $\langle Af|g\rangle=\langle f|A^Tg\rangle$ for all $A\in Sp(2n)$.
\end{proof}

\subsection{Orthogonality of the spherical Hermite polynomials}

Now choose a (real) orthogonal basis $\{ H_k^{(l)}\}$, $l =1, \ldots, \dim \cH_{k}^{f}$ of $\cH_k^f$ such that
\begin{equation}
\label{OrthSphHarm}
\langle H_k^{(p)} \exp(-\theta^2/2)|H_k^{(q)} \exp(-\theta^2/2) \rangle = \frac{\delta_{pq}}{(n-k)!}.
\end{equation}
A straightforward calculation shows that this is equivalent with
\begin{equation}
\label{OrthSphHarm2}
\langle H_k^{(p)} |H_k^{(q)} \rangle = \frac{\delta_{pq}}{2^{n-k}(n-k)!}.
\end{equation}
Using this basis, we consider the spherical Hermite functions $\varphi_{j,k,l}^{f}$ defined in (\ref{CliffordHermiteFunctions}). These functions are eigenfunctions of the harmonic oscillator (see (\ref{HOCH})), which immediately implies
\begin{equation*}
\langle \varphi_{j,k,l}^{f} |\varphi_{p,q,r}^{f} \rangle = 0
\end{equation*}
whenever $2j + k \neq 2p+q$. This can be generalized as follows.

\begin{theorem}
\label{orthocliffherm}
The spherical Hermite functions defined in equation (\ref{CliffordHermiteFunctions}), using a basis of fermionic harmonics satisfying (\ref{OrthSphHarm}), are orthogonal with respect to the inner product $\langle .|.\rangle $ on $\Lambda_{2n}$:
\begin{eqnarray*}
\langle \varphi_{j,k,l}^{f} |\varphi_{p,q,r}^{f} \rangle& =& \delta_{jp}\delta_{kq}\delta_{lr}\frac{4^{2j}j!}{(n-k-j)!}.
\end{eqnarray*}
\end{theorem}

\begin{proof}
Without loss of generality we assume  $j\ge p$. Using equation \eqref{CliffordHermiteFunctions}, corollary \ref{hermtoeg} and equation (\ref{differentiaalvgl}) and  we obtain
\begin{eqnarray*}
&&\langle \varphi_{j,k,l}^{f}|\varphi_{p,q,r}^{f} \rangle \\
&=& \langle H_k^{(l)}\exp (-\theta^2/2) | (\nabla^2_f+\theta^2 + 2\mE-2n)^j \varphi_{p,q,r}^{f}\rangle\\
&=&\delta_{jp}\frac{4^{2j}j!(n-q)!}{(n-q-j)!}\langle H_k^{(l)}\exp (-\theta^2/2) | H_q^{(r)}\exp (-\theta^2/2) \rangle.
\end{eqnarray*}
Because $H_k^{(l)}\exp (-\theta^2/2)$ and $H_q^{(r)}\exp (-\theta^2/2)$ belong to a different eigenspace of $H$ if $k\not=q$ and because of (\ref{OrthSphHarm}), we get the desired result. 
 \end{proof}

Denoting the normalization constants $\gamma^f_{j,k} = \langle \varphi_{j,k,l}^{f} |\varphi_{j,k,l}^{f} \rangle=\frac{4^{2j} j!}{(n-k-j)!}$ we have that the set of functions $\phi_{j,k,l}^{f} = \varphi_{j,k,l}^{f} / \sqrt{\gamma^f_{j,k}}$ satisfies
\[
\langle \phi_{j,k,l}^{f} |\phi_{p,q,r}^{f} \rangle = \delta_{jp} \delta_{kq} \delta_{lr}.
\]
Finally, we will need the following corollary.
\begin{corollary}
\label{orthsphharmferm}
For $H_k\in\cH_k^f$ and $H_l\in\cH_l^f$ with $k\not=l$ and $p(\theta^2)$ and $q(\theta^2)$ polynomials in $\theta^2$, the following holds
\[
\langle p(\theta^2) H_k|q(\theta^2)H_l\rangle =0.
\]
\end{corollary}
\begin{proof}
Since $p(\theta^2)H_k\in\mbox{span} \{\varphi^f_{j,k,l}| j\le n-k,l\le\dim\cH_k^f\}$, theorem \ref{orthocliffherm} implies the corollary.
 \end{proof}

\subsection{Mehler formula}

Now we have all tools necessary to obtain a Mehler formula in the Grassmann algebra $\Lambda_{2n}$. Let $f(\uxb)$ be an element of $\Lambda_{2n}$. Then it can be decomposed as $f(\uxb) = \sum_{j,k,l} a_{j,k,l} \phi_{j,k,l}^{f}(\uxb)$ with $a_{j,k,l} = \langle f |\phi_{j,k,l}^{f}\rangle$. We calculate the fractional Fourier transform of $f$, defined in formula (\ref{fracFourbasis}), as
\begin{eqnarray*}
\cF^{\alpha}_{0|2n} (f) &=& \sum_{j,k,l} a_{j,k,l} e^{i \alpha (2j+k)} \phi_{j,k,l}^{f}(\uyb)\\
&=& \sum_{j,k,l} \langle f |\phi_{j,k,l}\rangle e^{i \alpha (2j+k)} \phi_{j,k,l}^{f}(\uyb)\\
&=&\int_{B,x} f(\uxb) \sum_{j,k,l} \ast (\phi_{j,k,l}^{f})(\uxb) e^{i \alpha (2j+k)} \phi_{j,k,l}^{f}(\uyb)\\
&=&\int_{B,x}  \sum_{j,k,l} \ast (\phi_{j,k,l}^{f})(\uxb) e^{i \alpha (2j+k)} \phi_{j,k,l}^{f}(\uyb) f(\uxb).
\end{eqnarray*}
Comparing this expression with the integral representation of the fractional Fourier transform in theorem \ref{fracfourthm} (and using lemma \ref{triviaalstelemma}) yields the Mehler formula for Grassmann algebras:
\begin{eqnarray*}
&&\sum_{j,k,l} \ast (\phi_{j,k,l}^{f})(\uxb) e^{i \alpha (2j+k)} \phi_{j,k,l}^{f}(\uyb)\\
& =& \left(\pi (1- e^{2i \alpha})\right)^{n} \exp{\frac{  4 e^{i \alpha} \langle \uxb , \uyb \rangle -(1+ e^{2i \alpha})(\theta^2 + \theta_{\uyb}^2)}{2- 2e^{2i \alpha}}}
\end{eqnarray*}
with $k = 0, \ldots, n$; $j = 0, \ldots, n-k$ and $l = 1, \ldots, \dim \cH_{k}^{f}$. As this is a finite summation, there is no question of convergence. Using theorem \ref{CHstar} the left-hand side can be rewritten as:
\[
\sum_{j,k,l}\frac{(-1)^j}{\gamma^f_{j,k}} 4^{2j}(j!)^2 L_j^{(-n+k-1)}(\theta^2)\widetilde{H}_k^{(l)}(\uxb) e^{i \alpha (2j+k)}L_j^{(-n+k-1)}(\theta_{\uyb}^2)H_k^{(l)}(\uyb)\exp(-\frac{\theta^2+\theta_{\uyb}^2}{2}).
\]
In this expression there is a summation $\sum_l \widetilde{H}_k^{(l)}(\uxb) H_k^{(l)}(\uyb)$. This summation can be interpreted as a reproducing kernel for the space $\cH_{k}^{f}$. In the following theorem we obtain an explicit expression for this sum.

\begin{theorem}
\label{reprkern}
The function $F_k(\uxb,\uyb)= \sum_l \widetilde{H}_k^{(l)}(\uxb) H_k^{(l)}(\uyb)$ is given by
\begin{equation}
F_k(\uxb,\uyb)=\sum_{j=0}^{\lfloor \frac{k}{2}\rfloor}c^k_j (\langle \uxb , \uyb\rangle)^{k-2j}\theta^{2j}\theta_{\uyb}^{2j}
\label{reprkernfermformula}
\end{equation}
with constants $c_j^k=2^{k-2j}\pi^n\frac{(n+1-k)}{(k-2j)!j!(n+1+j-k)!}$.
\end{theorem}

\begin{proof}
We consider an element $\alpha(\uxb,\uyb)$ of the Grassmann algebra $\Lambda_{4n}$ generated by $\{{x\grave{}}_{1},{x\grave{}}_{2},\cdots,{x\grave{}}_{2n},{y\grave{}}_{1},\cdots,{y\grave{}}_{2n}\}$ which is harmonic, homogeneous of degree $k$ in both $\uxb$ and $\uyb$ and which satisfies
\begin{equation}
\label{fermreprkernelprop}
\langle\alpha(\uxb,\uyb)|H_k^{(r)}(\uyb)\rangle_{\uyb}=\frac{\widetilde{H}_k^{(r)}(\uxb)}{2^{n-k}(n-k)!}.
\end{equation}
The harmonicity implies that $\alpha$ has to be of the from $\sum_{l,t} c_{lt} \widetilde{H}_k^{(l)}(\uxb) H_k^{(t)}(\uyb)$ for some constants $c_{lt}$. Formula (\ref{fermreprkernelprop}) then implies $c_{lt}=\delta_{lt}$, so $\alpha$ is unique. This means $F_k$ is uniquely determined by these properties. 

On the other hand, for any $R_k\in\Lambda_{2n}(\uyb)$ of degree $k$ we calculate, using lemma \ref{adjoints},
\begin{eqnarray*}
\langle \langle \uxb,\uyb\rangle^k|R_k(\uyb)\rangle_{\uyb}&=&\frac{1}{2}\sum_{j=1}^{2n}\widetilde{x\grave{}}_j\langle \langle \uxb,\uyb\rangle^{k-1}{{y\grave{}}_j}|R_k(\uyb)\rangle_{\uyb}\\
&=&\left(\frac{1}{2}\right)^k\sum_{j_1,\cdots,j_k=1}^{2n}\widetilde{x\grave{}}_{j_k}\cdots\widetilde{x\grave{}}_{j_1}\langle {{y\grave{}}_{j_1}}\cdots {{y\grave{}}_{j_k}}|R_k(\uyb)\rangle_{\uyb}\\
&=&\sum_{j_1,\cdots,j_k=1}^{2n}\widetilde{{x\grave{}}_{j_1}\cdots {x\grave{}}_{j_k}}\langle1|\partial_{{y\grave{}}_{j_k}}\cdots\partial_{{y\grave{}}_{j_1}}R_k(\uyb)\rangle_{\uyb}\\
&=&\frac{k!}{(2\pi)^n}\widetilde{R_k(\uxb)}.
\end{eqnarray*}
This means that the normalized harmonic part of the Fischer decomposition of $\langle \uxb,\uyb\rangle^k$  given by equation (\ref{projectie}) will satisfy the conditions which uniquely define $F_k$. We hence conclude that 
\[
F_k(\uxb,\uyb)=\mP_0^k\left(\frac{(2\pi)^n}{k!2^{n-k}(n-k)!}\langle \uxb,\uyb\rangle^k\right).
\]
This can be calculated using the explicit form of the projection operators in equation (\ref{projectie}) and the fact that $\nabla_f^{2j}  \langle \uxb,\uyb\rangle^k = \frac{k!}{(k-2j)!}\langle \uxb,\uyb\rangle^{k-2j}\theta_{\uyb}^{2j}$, yielding
\begin{eqnarray*}
&&\mP_0^k\left(\frac{(2\pi)^n}{k!2^{n-k}(n-k)!}\langle \uxb,\uyb\rangle^k\right)\\
&=&\sum_{j=0}^{\lfloor \frac{k}{2}\rfloor}\frac{(2\pi)^n}{k!2^{n-k}(n-k)!}\frac{(n-k+1)!}{4^jj!(n-k+1+j)!}\frac{k!}{(k-2j)!}\theta^{2j}\theta_{\uyb}^{2j}\langle \uxb,\uyb\rangle^{k-2j}\\
&=&\sum_{j=0}^{\lfloor \frac{k}{2}\rfloor}2^{k-2j}\pi^n\frac{(n+1-k)}{(k-2j)!j!(n+1+j-k)!} \langle \uxb , \uyb\rangle^{k-2j}\theta^{2j}\theta_{\uyb}^{2j},
\end{eqnarray*}
which is the proposed formula.
 \end{proof}

The formula for $F_{k}$ given in theorem \ref{reprkern} can be seen as a dimensional continuation of the purely bosonic case in formula (\ref{bosreprkern}). The quotient of Gamma functions in the explicit expression of the Gegenbauer polynomials in (\ref{GegenbauerCoeffs}) (see Appendix) can be replaced by a Pochhammer symbol. This allows to define the Gegenbauer polynomials for $\alpha<-1/2$. Inspired by equation \eqref{bosreprkern} we start from $C^{(-n-1)}(t)$ and calculate
\begin{eqnarray*}
C^{(-n-1)}_{k}(t) &=& \sum_{j=0}^{\lfloor k/2\rfloor}\frac{(-1)^j(-n-1)_{k-j}}{j!(k-2j)!}(2t)^{k-2j}\\
&=&\sum_{j=0}^{\lfloor k/2\rfloor}\frac{(-1)^j(-n-1)(-n)\cdots(-n-2+k-j)}{j!(k-2j)!}(2t)^{k-2j}\\
&=&\sum_{j=0}^{\lfloor k/2\rfloor}\frac{(-1)^k(n+1)(n)\cdots(n+2-k+j)}{j!(k-2j)!}(2t)^{k-2j}\\
&=&(-1)^k(n+1)!\sum_{j=0}^{\lfloor k/2\rfloor}\frac{1}{(k-2j)!j!(n+1+j-k)!}(2t)^{k-2j}.
\end{eqnarray*}
Comparison with (\ref{reprkernfermformula}) then gives
\[
F_k(\uxb,\uyb)= \pi^{n}(-1)^k\frac{n+1-k}{(n+1)!} \left(\sqrt{\theta^{2}\theta_{\uyb}^{2}}\right)^{k} C^{(-n-1)}_k\left(\frac{\langle \uxb , \uyb\rangle}{\sqrt{\theta^{2}\theta_{\uyb}^{2}} }\right). 
\]

\begin{remark}
It is interesting to note that the reproducing kernel for spaces of harmonics is always expressed by Gegenbauer polynomials, in the case of orthogonal symmetry, finite reflection group symmetry as well as symplectic symmetry.
\end{remark}

Using the expression for $F_k$, we obtain the following simplification of the Mehler formula
\begin{eqnarray*}
&&\sum_{k=1}^n\sum_{j=1}^{n-k}(-1)^jj!(n-k-j)! L_j^{k-n-1}(\theta^2) e^{i \alpha (2j+k)}L_j^{k-n-1}(\theta_{\uyb}^2 ) F_k(\uxb,\uyb)\\
&=& \left(\pi (1- e^{2i \alpha})\right)^{n} \exp{\frac{ 2 e^{i \alpha} \langle \uxb , \uyb \rangle - e^{2i \alpha}(\theta^2 +\theta_{\uyb}^2)}{1- e^{2i \alpha}}}.
\end{eqnarray*}
In the limit case $\alpha = \pm \pi/2$ (corresponding with the classical Fourier transform) this formula reduces to
\begin{eqnarray*}
&&\sum_{j,k}j!(n-k-j)! L_j^{k-n-1}(\theta^2) (\pm i)^kL_j^{k-n-1}(\theta_{\uyb}^2 ) F_k(\uxb,\uyb) \exp(-\frac{\theta^2+\theta_{\uyb}^2}{2})\\
 &&= \left(2\pi \right)^{n} \exp{ \pm i \langle \uxb , \uyb \rangle}.
\end{eqnarray*}

\begin{remark}
By making use of the expression for the reproducing kernel, we immediately see that both the left-hand and right-hand side in the Mehler formula are invariant under the symplectic group $Sp(2n)$, acting simultaneously on $\uxb$ and $\uyb$. So we have indeed constructed a symplectic analog of the $O(m)$-invariant Mehler formula.
\end{remark}

\section{Hermite polynomials for $O(m)\times Sp(2n)$ symmetry}
\setcounter{equation}{0}
\label{supersuper}

\subsection{Orthogonality of spherical Hermite polynomials and inner products}
\label{innerprodsSuper}

We start by introducing the canonical inner product on $L_2(\mR^m)\otimes \Lambda_{2n}$ by combining the standard bosonic $L_2$ inner product and the fermionic inner product from definition \ref{inprodGrass}.
\begin{definition}
\label{Canonischin}
The inner product $\langle .|. \rangle_1: L_2(\mR^{m})_{m|2n}\times L_2(\mR^{m})_{m|2n}\to \mC$ is given by
\[
\langle f | g\rangle_1 = \int_{\mR^{m|2n}} f (* \overline{g})
\]
where the star map acts on $\Lambda_{2n}$ as in definition \ref{defstar} and leaves the bosonic variables invariant.
\end{definition}

This inner product is uniquely determined by demanding that the creation and annihilation operators of the harmonic oscillator are mutually adjoint. This is the subject of the following theorem.

\begin{theorem}
The inner product $\langle . | . \rangle_1$ is, up to a multiplicative constant, the unique inner product on $\cP \exp(-R^2/2)$ such that $(a_i^\pm)^{\dagger}=a_i^\mp$ and $(b_j^\pm)^{\dagger}=b_j^\mp$, with $a_j$ and $b_j$ the creation and annihilation operators appearing in the hamiltonian of the harmonic oscillator (\ref{hamiltoniaan}).
\end{theorem}

\begin{proof}
The condition $(a_i^\pm)^{\dagger}=a_i^\mp$ is equivalent to $x_j^\dagger=x_j$ and $\partial_{x_j}^\dagger=-\partial_{x_j}$. Demanding that $(b_j^{\pm})^{\dagger}=b_j^\mp$ for every $j$ is equivalent to demanding $\partial_{{x \grave{}}_{j}}^\dagger={x \grave{}}_{j}/2$. These conditions are clearly fulfilled for $\langle . | . \rangle_1$. Because these conditions determine the fermionic inner product (lemma \ref{adjoints}) and the bosonic inner product completely this also holds for the full inner product $\langle . | . \rangle_1$.
 \end{proof}

As a consequence of this theorem, it is easy to compute that the set of functions 
\begin{eqnarray*}
\psi_{k_{1}, \ldots, k_{m}; l_{1}, \ldots, l_{2n}}&=& \frac{1}{\sqrt{  k_{1}! \ldots k_{m}! \pi^{M/2}}}\\
&&\times (a_1^+)^{k_1}(a_2^+)^{k_2}\ldots (a_m^+)^{k_m} (b_1^+)^{l_1}\ldots (b_{2n}^+)^{l_{2n}}\exp (-R^{2}/2),
\end{eqnarray*}
with $k_{i} \in \mN$ and $l_{j} \in \{0,1\}$ is an orthonormal basis of  $\cP \exp(-R^2/2)$, i.e. 
\[
\langle \psi_{k_{1}, \ldots, k_{m}; l_{1}, \ldots, l_{2n}}, \psi_{p_{1}, \ldots, p_{m}; q_{1}, \ldots, q_{2n}}\rangle_{1} = \delta_{k_{1} p_{1}} \ldots \delta_{k_{m} p_{m}} \delta_{l_{1} q_{1}} \ldots \delta_{l_{2n} q_{2n}}. 
\]
This basis should be considered as the super-analog of the cartesian basis $\{\psi_{k_{1}, \ldots, k_{m}}^{b}\}$ introduced in the case of $O(m)$ symmetry.

The inner product $\langle . | . \rangle_1$ also has several undesirable properties, namely
\begin{itemize}
\item $\nabla^2$ and $R^{2}$ are neither self-adjoint nor mutually adjoint
\item the spherical Hermite functions are in general not orthogonal
\item contrary to the purely fermionic case, the star map does not preserve harmonicity.
\end{itemize}

Indeed, using corollary \ref{hermtoeg} we find $(R^2)^\dagger=r^2-\nabla^2_f$ and $(\nabla^2)^\dagger=\nabla^2_b-\theta^2$. This means for instance that
\begin{equation}
\label{toegDelta}
(-\nabla^2 -R^2 +2\mE +M)^\dagger =(-\nabla^2_b -r^2+\theta^2+\nabla^2_f -2\mE_b -m+2\mE_f -2n).
\end{equation}
This implies that the procedure, used in theorem \ref{orthocliffherm} to prove the orthogonality of the spherical Hermite functions, is no longer possible. Moreover, an easy example shows that the spherical Hermite functions are indeed not orthogonal.

\begin{example}
We take the spherical harmonics $H_2=2x_1^2-{x\grave{}}_1{x\grave{}}_2$, $H_0=1$ and form $L_1^{\frac{M}{2}-1}(R^2)H_0=-R^2+M/2$. By using equation (\ref{CHexp}) for the purely bosonic and fermionic case we find
\begin{equation*}
(-\nabla^2_b -r^2+\theta^2+\nabla^2_f -2\mE_b -m+2\mE_f -2n)\exp(-R^2/2)=\exp(-R^2/2)(-\nabla^2_b+\nabla^2_f ).
\end{equation*}
Using this and equations \eqref{CliffordHermiteFunctionssuper}, (\ref{toegDelta}) and equation \eqref{CHexp} we obtain
\begin{eqnarray*}
&&\langle H_2\exp (-R^2/2) | 4L_1^{\frac{M}{2}-1}(R^2)H_0\exp (-R^2/2)\rangle_1 \\
 &=& \langle H_2\exp (-R^2/2) | (-\nabla^2 -R^2+2\mE+M)\exp (-R^2/2)\rangle_1\\
&=&\langle [(-\nabla^2_b+\nabla^2_f) H_2]\exp (-R^2/2) |\exp (-R^2/2)\rangle_1\\
&=&\langle -8\exp(-R^2/2)|\exp (-R^2/2)\rangle_1.
\end{eqnarray*}
Since $\langle \exp(-R^2/2)|\exp (-R^2/2)\rangle_1 \neq 0$, these functions are not orthogonal.
\end{example}

However, it is still true that the hamiltonian $H$ of the harmonic oscillator is hermitian. The spherical Hermite functions are eigenvectors of this operator, see (\ref{HOCH}). Hence, the spherical Hermite functions belonging to different eigenvalues are still orthogonal. In the example, we see that both $H_2\exp(-R^2/2)$ and $L_1^{\frac{M}{2}-1}(R^2)H_0\exp(-R^2/2)$ have eigenvalue $M/2+2$. 

We can make this partial orthogonality even stronger. 
\begin{lemma}
\label{bijnaorth}
With $\{ H_{k}^{b(l)} \}$ the basis in formula \eqref{bosharmbasis} for $\cH_k^b$ and $\{ H_p^{f(q)} \}$ the basis in formula \eqref{OrthSphHarm} for $\cH_{p}^{f}$, one has
\[
\langle L_j^{\frac{M}{2}+2i+k+p-1}(R^2)f_{i,k,p}H_{k}^{b(l)}H_p^{f(q)}\exp(-R^2/2)|  L_s^{\frac{M}{2}+2u+t+v-1}(R^2)f_{u,t,v}H_{t}^{b(w)}H_v^{f(z)}\exp(-R^2/2) \rangle_{1} = 0
\]
unless $k=t$, $p=v$, $l=w$, $q=z$ and $j+i=s+u$.
\end{lemma}

\begin{proof}
Recall that the inner product $\langle . |.  \rangle_{1}$ is a combination of the known bosonic and fermionic inner product. Hence, using 
corollary \ref{orthsphharmferm} and the orthogonality of (bosonic) spherical harmonics over the unit sphere, it follows that the integral is zero unless $k=t$, $p=v$, $l=w$ and $q=z$. The self-adjointness of the hamiltonian of the harmonic oscillator then implies $2j+2i+k+p=2s+2u+t+v$, from which we obtain $j+i=s+u$.
 \end{proof}

As a corollary we obtain the following decomposition of $\cP \exp(-R^2/2)$. 
\begin{corollary}
With respect to $\langle\cdot|\cdot\rangle_1$, the space $\cP \exp(-R^2/2)$ decomposes in mutually orthogonal subspaces of dimension at most $n+1$, spanned by the spherical Hermite functions.
\end{corollary}

\begin{proof}
This follows from  lemma \ref{bijnaorth} together with the fact that for given $H_k^b\in\cH_k^b$ and $H_p^{f} \in \cH_{p}^{f}$ there are at most $n+1$ polynomials $f_{i,k,p}$ (see theorem \ref{decompintoirreps}).
 \end{proof}

Finally, another problem with the inner product $\langle . |.  \rangle_{1}$ is that, contrary to the purely fermionic case (see lemma \ref{expstar} $(ii)$), $* H_k \exp (-R^2/2)$ with $H_{k} \in \cH_{k}$ will not always be an element of $\cH_k \exp (-R^2/2)$. This problem occurs for spherical harmonics of the form $f_{k,p,q}H_p^bH_q^f$ with $k>0$ (see theorem \ref{decompintoirreps}). It is however possible to introduce a new inner product on $\cP\exp(-R^2/2)$, which makes the spherical Hermite functions orthogonal and also solves the other problems.

From now on, we will always assume  $M>0$. At the end of the section we will explicitly show that the following constructions are not possible for $M\le 0$ (see theorem \ref{Nogonegdim}). The assumption $M>0$ means that we have a Fischer decomposition $(\ref{superFischer})$ and that the spherical Hermite polynomials constitute a basis for $\cP$.

The orthogonality of the bosonic spherical Hermite polynomials depends on two facts. The spherical harmonics are orthogonal with respect to integration over the unit sphere and the radial part (given by Laguerre polynomials) is orthogonal with respect to the radial integration. Therefore, our first aim is to construct an inner product for the spherical harmonics using integration over the supersphere. We start with a few technical lemmas.

\begin{lemma}
\label{SSin1}
For the polynomials $f_{k,p,q}$ introduced in lemma \ref{polythm}, $H_p^b \in \cH_p^b$ and $H_q^f \in \cH_q^f$, the following algebraic relation holds
\begin{eqnarray*}
f_{k,p,q}(r^2,\theta^2)H_p^bH_q^f&\equiv&(-1)^ka_{k,p,q}r^{2k}H_p^bH_q^f\mod R^2
\end{eqnarray*}
with $a_{k,p,q}=\frac{\Gamma(M/2+p+q+2k-1)}{\Gamma(M/2+p+q+k-1)}$.
\end{lemma}

\begin{proof}
To calculate $a_{k,p,q}$ explicitly we start from
\[
f_{k,p,q}H_p^bH_q^f \equiv \sum_{s=0}^ka_sr^{2k-2s}(-r^2)^{s}H_p^bH_q^f\mod R^2
\]
which leads to
\[
a_{k,p,q}=\frac{\Gamma(\frac{m}{2}+p+k)}{(n-q-k)!}\sum_{s=0}^k(-1)^{k-s}\binom{k}{s}\frac{(n-q-s)!}{\Gamma(\frac{m}{2}+p+k-s)}
\]
and the result follows from lemma \ref{SSin2}.
 \end{proof}

\begin{lemma}
\label{SSin2}
For $\nu\in\mN$, $\mu\in\mR$ and $\mu>\nu$ the following relation holds
\begin{eqnarray*}
\sum_{s=0}^k(-1)^{k-s}\binom{k}{s}\frac{(\nu-s)!}{\Gamma(\mu+k-s)}&=&\frac{\Gamma(\mu-\nu+2k-1)}{\Gamma(\mu-\nu+k-1)}\frac{(\nu-k)!}{\Gamma(\mu+k)}.
\end{eqnarray*}
\end{lemma}
\begin{proof}
We denote $c^k_{\mu,\nu}=\sum_{s=0}^k(-1)^{k-s}\binom{k}{s}\frac{(\nu-s)!}{\Gamma(\mu+k-s)}$. Then $c_{\mu,\nu}^0=\nu!/\Gamma(\mu)$ holds and for $k=1$ we find
\begin{eqnarray*}
c^1_{\mu,\nu}&=&-\frac{\nu!}{\Gamma(\mu+1)}+\frac{(\nu-1)!}{\Gamma(\mu)}=\frac{(\nu-1)!}{\Gamma(\mu+1)}\left(\mu-\nu \right)\\
&=&\frac{\Gamma(\mu-\nu+1)}{\Gamma(\mu-\nu)}\frac{(\nu-1)!}{\Gamma(\mu+1)}.
\end{eqnarray*}
From the definition of $c_{\mu,\nu}^k$ we calculate
\begin{eqnarray*}
c_{\mu,\nu}^k&=&\sum_{s=0}^{k-1}(-1)^{k-s}\binom{k-1}{s}\frac{(\nu-s)!}{\Gamma(\mu+k-s)}+\sum_{s=1}^k(-1)^{k-s}\binom{k-1}{s-1}\frac{(\nu-s)!}{\Gamma(\mu+k-s)}\\
&=&\sum_{s=0}^{k-1}(-1)^{k-s}\binom{k-1}{s}\frac{(\nu-s)!}{\Gamma((\mu+1)+k-1-s)}\\
&& -\sum_{s=0}^{k-1}(-1)^{k-s}\binom{k-1}{s}\frac{(\nu-s-1)!}{\Gamma(\mu+k-s-1)}\\
&=&-c_{\mu+1,\nu}^{k-1}+c^{k-1}_{\mu,\nu-1}.
\end{eqnarray*}
Since $\mu-\nu+1>0$, this implies that, if the lemma holds for $k-1$, then
\begin{eqnarray*}
c_{\mu,\nu}^k&=&-\frac{\Gamma(\mu-\nu+2k-2)}{\Gamma(\mu-\nu+k-1)}\frac{(\nu-k+1)!}{\Gamma(\mu+k)}\\
&& +\frac{\Gamma(\mu-\nu+2k-2)}{\Gamma(\mu-\nu+k-1)}\frac{(\nu-k)!}{\Gamma(\mu+k-1)}\\
&=&\frac{\Gamma(\mu-\nu+2k-2)}{\Gamma(\mu-\nu+k-1)}\frac{(\nu-k)!}{\Gamma(\mu+k)}\left(\mu+k-1-(\nu-k+1)\right)\\
&=&\frac{\Gamma(\mu-\nu+2k-1)}{\Gamma(\mu-\nu+k-1)}\frac{(\nu-k)!}{\Gamma(\mu+k)},
\end{eqnarray*}
so the lemma is proven by induction.
 \end{proof}

We will also need the following result.

\begin{lemma}
\label{SSin3}
For every $k\in\mN$ and $\alpha\in\mR$
\[
\sum_{s=0}^k(-1)^s\binom{k}{s}(\alpha-s)_{k}=k!
\]
with $(a)_k=a(a+1)\cdots(a+k-1)$ the Pochhammer symbol.
\end{lemma}
\begin{proof}
We prove this lemma again by induction. For $k=1$ the lemma is trivial. We will use lemma 5 in \cite{DBThesis} which states
\begin{eqnarray}
\sum_{s=0}^k(-1)^s\binom{k}{s}(\alpha-s)_{k-1}=0.
\label{lemThesis}
\end{eqnarray}
From the definition we calculate, using (\ref{lemThesis}),
\begin{eqnarray*}
\sum_{s=0}^k(-1)^s\binom{k}{s}(\alpha-s)_{k}&=& \sum_{s=0}^k(-1)^s\binom{k}{s}(\alpha-s+k-1)(\alpha-s)_{k-1}\\
&=&- \sum_{s=1}^k(-1)^ss\binom{k}{s}(\alpha-s)_{k-1}\\
&=&\sum_{s=0}^{k-1}(-1)^s\frac{(s+1) k!}{(k-s-1)!(s+1)!}(\alpha-s-1)_{k-1}\\
&=&k\sum_{s=0}^{k-1}(-1)^s\binom{k-1}{s}((\alpha-1)-s)_{k-1},
\end{eqnarray*}
so the lemma follows by induction.
 \end{proof}

Now we have all the necessary tools to prove the following lemma.

\begin{lemma}
\label{SSin4}
For $\{H_p^{b(l)}\}$ the orthonormal basis of $\cH_{p}^{b}$ in formula \eqref{bosharmbasis}, $\{H_q^{f(t)}\}$ the orthonormal basis of $\cH_{q}^{f}$ in equation \eqref{OrthSphHarm} and $f_{k,p,q}$ as defined in lemma \ref{polythm}, the following relation holds
\begin{eqnarray*}
&&\int_{SS}f_{k,p,q}H_p^{b(l_1)}H_q^{f(t_1)}\, f_{k,p,q}H_p^{b(l_2)}\widetilde{H}_q^{f(t_2)}\\
&=&(-1)^k\frac{ k!\,\Gamma(\frac{m}{2}+p+k)}{(\frac{M}{2}+p+q+2k-1)\Gamma(\frac{M}{2}+p+q+k-1)(n-q-k)!}\delta_{l_1l_2}\delta_{t_1t_2}.
\end{eqnarray*}
\end{lemma}

\begin{proof}
Lemma \ref{SSin1} implies there exists a polynomial $P$ of degree $2k+p+q-2$, such that $f_{k,p,q}H_p^bH_q^f=(-1)^ka_{k,p,q} r^{2k}H_p^bH_q^f+R^2 P$. Since $M>0$, $P$ has a Fischer decomposition containing only spherical harmonics of degree smaller than or equal to $2k+p+q-2$. The orthogonality of spherical harmonics of different degree and the property $\int_{SS}R^2f=\int_{SS}f$ then imply
\begin{eqnarray*}
&&\int_{SS}f_{k,p,q}H_p^{b(l_1)}H_q^{f(t_1)}\, f_{k,p,q}H_p^{b(l_2)}\widetilde{H}_q^{f(t_2)}\\
&=&(-1)^ka_{k,p,q}\int_{SS}r^{2k} f_{k,p,q}H_p^{b(l_1)}H_p^{b(l_2)} H_q^{f(t_1)}\widetilde{H}_q^{f(t_2)}\\
&=&(-1)^ka_{k,p,q}\sum_{s=0}^ka_s\int_{SS}r^{4k-2s} \theta^{2s}H_p^{b(l_1)}H_p^{b(l_2)} H_q^{f(t_1)}\widetilde{H}_q^{f(t_2)}.
\end{eqnarray*}
Using expression \eqref{SSintnieuw} for the supersphere integration, we then obtain
\begin{eqnarray*}
&=&(-1)^ka_{k,p,q}\sum_{s=0}^ka_s\int_{\mS^{m-1}} d \sigma \int_B\frac{(-1)^{n-q-s}\theta^{2n-2q-2s}}{(n-q-s)!} \\
&&\times \left[\left(\frac{\partial}{\partial r^2}\right)^{n-q-s}r^{4k-2s+m-2} \theta^{2s}H_p^{b(l_1)}H_p^{b(l_2)} H_q^{f(t_1)}\widetilde{H}_q^{f(t_2)}\right]_{r=1}\\
&=&(-1)^ka_{k,p,q}\sum_{s=0}^k(-1)^sa_s\delta_{l_1l_2}\frac{\delta_{t_1t_2}}{(n-q-s)!}\left[\left(\frac{\partial}{\partial r^2}\right)^{n-q-s}r^{2(2k-s+m/2+p-1)}\right]_{r=1}\\
&=&(-1)^k\delta_{l_1l_2}\delta_{t_1t_2}a_{k,p,q}\sum_{s=0}^k(-1)^s\frac{a_s}{{(n-q-s)!}}\frac{\Gamma(2k-s+\frac{m}{2}+p)}{\Gamma(2k-s+\frac{m}{2}+p-n+q+s)}\\
&=&(-1)^k\delta_{l_1l_2}\delta_{t_1t_2}\frac{a_{k,p,q}}{{\Gamma(2k+\frac{M}{2}+p+q)}}\frac{\Gamma(\frac{m}{2}+p+k)}{(n-q-k)!}\\
&&\times \sum_{s=0}^k(-1)^s\binom{k}{s}\frac{\Gamma(2k-s+\frac{m}{2}+p)}{\Gamma(\frac{m}{2}+p+k-s)}\\
&=&(-1)^k\delta_{l_1l_2}\delta_{t_1t_2}\,a_{k,p,q}\,b_{k,p,q}.
\end{eqnarray*}
Lemma \ref{SSin3} implies
\begin{equation}
\label{bkpq}
b_{k,p,q}=\frac{k!}{{\Gamma(2k+\frac{M}{2}+p+q)}}\frac{\Gamma(\frac{m}{2}+p+k)}{(n-q-k)!}.
\end{equation}
Substituting $a_{k,p,q}$ from lemma \ref{SSin1} yields the desired result.
 \end{proof}

Now we introduce the following isomorphism $T: \cH_{k} \rightarrow \cH_{k}$ given by
\[
T(H_k)=(-1)^{i}f_{i,k-2i-j,j}(r^2,\theta^2) H^b_{k-2i-j}\widetilde{H}^f_j,
\]
for $H_{k} = f_{i,k-2i-j,j}(r^2,\theta^2) H^b_{k-2i-j}H^f_j$ with $H^b_{k-2i-j} \in \cH^b_{k-2i-j}$, $H^f_j \in \cH^f_j$  and extended by linearity to the whole of $\cH_{k}$. 
\begin{remark}
\label{TOSp}
Using lemma \ref{Sptilde} it is straightforward to prove that for $A\in O(m)\times Sp(2n)$
\[
T\circ A = (G^TAG)\circ T
\]
holds, with $G$ given in equation \eqref{metricG}.
\end{remark}

Using the isomorphism $T$ we can construct a new inner product on $\bigoplus_{k=0}^{\infty} \cH_k \exp (-R^2/2)$. This is the subject of the following theorem.

\begin{theorem}
\label{defsuperH}
Put $\cH = \bigoplus_{k=0}^{\infty} \cH_k = \ker{\nabla^2} \, \cap \, \cP$.
The product $\langle .|.\rangle: \cH \exp (-R^2/2) \times \cH \exp (-R^2/2)\to \mC$ given by
\[
\langle H_1\exp( -R^2/2)| H_2\exp( -R^2/2)\rangle =\int_{\mR^{m|2n}}H_1\overline{T({H}_2)}\exp (-R^2)
\]
with $H_{1}, H_{2} \in \cH$ is an inner product on $\cH \exp (-R^2/2)$ satisfying
\[
\langle \cH_k \exp (-R^2/2) , \cH_l \exp (-R^2/2)\rangle = 0
\]
if $k\not= l$.
\end{theorem}

\begin{proof}
The product is clearly linear. It is positive definite since 
\[
\langle\left(f_{i,p,q} \cH^b_{p} \otimes \cH^f_{q}\right) \exp (-R^2/2) ,\left(f_{j,r,s} \cH^b_{r} \otimes \cH^f_{s}\right) \exp (-R^2/2)\rangle = 0
\]
when $(i,p,q) \neq (j,r,s)$ (see theorem \ref{integorth}) and $\langle f_{i,p,q}H_p^bH_q^f|f_{i,p,q}H_p^bH_q^f\rangle>0$ (see lemma \ref{SSin4}). Moreover, lemma \ref{SSin4} also implies the inner product is symmetric.
 \end{proof}

This inner product can now be extended to the whole space using decomposition (\ref{superFischer}).

\begin{theorem}
\label{defsuper}
The product $\langle .|. \rangle_{2}: \cP \exp (-R^2/2)\times \cP \exp (-R^2/2)\to \mC$ given by
\[
\langle R^{2i}H_k\exp (-R^2/2) | R^{2j}H_l\exp (-R^2/2)\rangle_2 = \int_{\mR^{m|2n}} R^{2i+2j} H_k \, \overline{T({H}_l)} \, \exp (-R^2)
\]
with $H_{k} \in \cH_{k}$, $H_{l} \in \cH_{l}$ and extended by linearity is an inner product.
\end{theorem}

\begin{proof}
The product is clearly linear and symmetric. Using (\ref{superint2}) and theorem \ref{defsuperH} we subsequently obtain 
\begin{eqnarray*}
&&\langle R^{2i}H_k\exp (-R^2/2) | R^{2j}H_k\exp (-R^2/2)\rangle_{2}\\
 &=&\frac{1}{2}\Gamma \left(\frac{i+j+2k+M}{2}\right)\int_{SS}H_k \overline{T({H}_l)}\\
 &=&\frac{\Gamma \left(\frac{i+j+2k+M}{2}\right)}{\Gamma \left(\frac{2k+M}{2}\right)} \int_{\mR^{m|2n}} H_k \, \overline{T({H}_l)} \, \exp (-R^2)\\
 &\geq&0.
\end{eqnarray*}
Hence $\langle.|. \rangle_{2}$ is positive definite and defines an inner product. 
 \end{proof}
 
The behavior of the inner product with respect to $O(m)\times Sp(2n)$ is given by 
\begin{lemma}
For $A\in O(m)\times Sp(2n)$ and for $f,g\in\cP\exp(-R^2/2)$, with action on superfunctions as defined in formula \ref{actieOSp}, the relation
\[
\langle Af|g\rangle=\langle f|A^Tg\rangle
\]
holds. This implies $A^\dagger=A^T$ for all $A\in O(m)\times Sp(2n)$ and is equivalent with
\[
\langle AG f|GAg\rangle_2=\langle  f|g\rangle_2,
\]
with $G$ given in equation \eqref{metricG}.
\end{lemma}
\begin{proof}
The proof is similar to the proof of proposition \ref{Spinprod}. Remark \ref{TOSp} and the $O(m)\times Sp(2n)$-invariance of $\int_{\mR^{m|2n}}$ (see lemma \ref{Berezincov}) lead to the proposed formula.
\end{proof}
The inner product of theorem \ref{defsuper} can be written more concisely as
\[
\langle f | g \rangle_{2} = \int_{\mR^{m|2n}} f \overline{T(g)}, \qquad f, g \in \cP \exp(-R^{2}/2)
\] 
by extending $T$ to $\cP\exp(-R^2/2)$ such that 
\begin{equation*}
T( R^{2j}H_l\exp (-R^2/2))=R^{2j}T({H}_l)\exp (-R^2/2), \quad H_{l} \in \cH_{l}.
\end{equation*}
So $T$ is now a map $\cP\exp(-R^2/2)\to\cP\exp(-R^2/2)$ satisfying
\begin{equation*}
T(L_j^{\frac{M}{2}+k-1}(R^2)H_k^{(l)}\exp(-R^2/2))= L_j^{\frac{M}{2}+k-1}(R^2)T({H}_k^{(l)})\exp(-R^2/2).
\end{equation*}
It is clear by construction that $T$ preserves harmonicity. Now we can show that $\langle .|.\rangle_2 $ also satisfies the other desirable properties that $\langle.|.\rangle_1$ did not satisfy. We start by proving that $R^2$ and $\nabla^2$ are self-adjoint.

\begin{lemma}
\label{adjointsnew}
One has 
\begin{eqnarray*}
(R^2)^{\dagger}&=&R^2\\
(\nabla^2)^{\dagger}&=&\nabla^2\\
(2\mE+M)^{\dagger}&=&-(2\mE+M)
\end{eqnarray*}
with respect to the inner product $\langle. |.\rangle_2$ on $\cP\exp (-R^2/2)$.
\end{lemma}

\begin{proof}
The proof of the first property is trivial. The second property is obtained as follows. Because of lemma \ref{Berezin} we know that 
\[
\int_{\mR^{m|2n}} f \nabla^2 g=\int_{\mR^{m|2n}} \nabla^2(f) g.
\]
Using lemma \ref{laplonpieces} and equation (\ref{berekexpdelta}) we obtain that for every piece of the Fischer decomposition
\begin{eqnarray*}
&&T(\nabla^2 R^{2j}H_k\exp(-R^2/2))\\
&=&T([c_{1,j,k} R^{2j-2}H_k+R^{2j+2}H_k+(4j+2k+M)R^{2j}H_k]\exp(-R^2/2))\\
&=&[c_{1,j,k} R^{2j-2}+R^{2j+2}+(4j+2k+M)R^{2j}]T(H_k)\exp(-R^2/2)\\
&=&\nabla^2 R^{2j}T(H_k)\exp(-R^2/2)\\
&=&\nabla^2 T( R^{2j} H_k\exp(-R^2/2)),
\end{eqnarray*}
with $H_{k} \in \cH_{k}$.
Combining these two results yields the second property.  
Finally, the result for $2\mE + M$ follows immediately from equation \eqref{commsuper}.
 \end{proof}

To show that the spherical Hermite functions are orthogonal, we first need an orthogonal basis of spherical harmonics. The knowledge of orthonormal bases for the bosonic and fermionic harmonic polynomials suffices to find an orthonormal basis $\{ H_{k}^{(l)}\}$ for the space $\cH_{k}$, satisfying
\begin{equation}
\label{orthbasissuper}
\int_{SS}H_k^{(l)}T({H}_q^{(r)})=\delta_{kq}\delta_{lr}.
\end{equation}
Indeed, using the bases in equations \eqref{bosharmbasis} and \eqref{OrthSphHarm}, it is easy to check that
\begin{eqnarray*}
&&\left\{\frac{f_{i,k-2i-j,j}H_{k-2i-j}^{b(l)}H_{j}^{f(t)}}{\sqrt{a_{i,k-2i-j,j} b_{i,k-2i-j,j}}}|0\le j\le \min(n,k-1)-1,\, \right. \\
 &&\left. 0\le i\le \min(n-j,\lfloor\frac{k-j}{2}\rfloor)\right\}
\end{eqnarray*}
with $l\le\dim\cH_{k-2i-j}^b$ and $t\le \dim\cH_j^f$ is an orthonormal basis for $\cH_k$.

For this basis of spherical harmonics $\{H_k^{(l)}\}$ the following theorem holds.

\begin{theorem}[Orthogonality spherical Hermite functions]
\label{orthsuperCH}
The set of functions $\{ \varphi_{j,k,l} \}$ in formula \eqref{CliffordHermiteFunctionssuper}, constructed using  the basis of spherical harmonics in formula \eqref{orthbasissuper}, forms an orthogonal basis for $\cP\, exp(-R^{2}/2)$ with respect to the inner product $\langle .|.\rangle_2$. The normalization is given by
\begin{equation}
\langle \varphi_{j,k,l}| \varphi_{p,q,r} \rangle_2 =\frac{1}{2}4^{2j}j!\Gamma (j+k + M/2)\delta_{jp}\delta_{kq}\delta_{lr}=\gamma^M_{j,k}\delta_{jp}\delta_{kq}\delta_{lr}.
\label{normCH}
\end{equation}
\end{theorem}

\begin{proof}
This result is proven with the same technique as in theorem \ref{orthocliffherm}, using the results we obtained in lemma \ref{adjointsnew}.
 \end{proof}

Note that the normalization constants only depend op the super-dimension $M$ and not on the bosonic and fermionic dimensions separately. In particular this implies that the normalization constants are equal to the case with $M$ bosonic variables (and no anti-commuting variables).

Both the inner products $\langle . | . \rangle_1$ and $\langle . | . \rangle_2$ have their own advantages. The hamiltonian of the harmonic oscillator $\frac{1}{2}(-\nabla^2+R^2)$ is hermitian for both. For the inner product $\langle . | . \rangle_1$ we even have $(a_i^{\pm})^\dagger=a_i^\mp$ and $(b_j^\pm)^\dagger=b_j^{\mp}$ or equivalently $x_i^\dagger=x_i$ and $\partial_{{x \grave{}}_{j}}^\dagger={x \grave{}}_{j}/2$. However, for the inner product $\langle . | . \rangle_2$ we have $(R^2)^\dagger= R^2$ and $(\nabla^2)^\dagger= \nabla^2$. This is of major importance to study other potentials in superspace, such as anharmonic oscillators. Using $\langle . | . \rangle_2$, they still have symmetric hamiltonians. That is why we will study the extension of $\langle.|.\rangle_2$ from $\cP \exp (-R^{2}/2)$ to the super Schwartz and $L_2$-space in a forthcoming article.

We end this section with two no-go results. First, it is not possible to construct an inner product which has the advantages of both inner products $\langle . | . \rangle_1$ and $\langle . | . \rangle_2$. This is the subject of the following theorem.
\begin{theorem}
\label{Nogo1}
There is no inner product on $\cP\exp(-R^2/2)$ for which $(R^2)^\dagger=R^2$ and $(r^2)^\dagger=r^2$. 
\end{theorem}
\begin{proof}
$(R^2)^\dagger=R^2$ and $(r^2)^\dagger=r^2$ imply that $(\theta^2)^\dagger=\theta^2$. This is impossible because of the same reason as in theorem \ref{onmogelijkinprod}. 
 \end{proof}

We have only found an inner product with the property that $(R^2)^\dagger=R^2$ and $(\nabla^2)^\dagger=\nabla^2$ for the case $M>0$. Now we will show that such an inner product does not exist in case $M\le 0$. We also prove that (for all $M$) there does not exist an inner product with the properties of the purely fermionic inner product (see corollary \ref{hermtoeg}) when $m\not=0$.

\begin{theorem}
\label{Nogonegdim}
There is no inner product on $\cP\exp (-R^2/2)$ for which $(R^2)^\dagger=R^2$ and $(\nabla^2)^\dagger=\nabla^2 $ in case $M\le 0$. 

There also does not exist an inner product with the properties $(R^2)^\dagger=-\nabla^2$ and $(\nabla^2)^\dagger=-R^2$ for arbitrary $M$ with $m\not=0$.
\end{theorem}

\begin{proof}
If $(R^2)^\dagger=R^2$ and $(\nabla^2)^\dagger=\nabla^2 $, using equation \eqref{commsuper} we obtain that $(2\mE+M)^\dagger=\frac{1}{2}([\nabla^2, R^2])^\dagger=-(2\mE+M)$. So assume there exists an inner product satisfying these properties for $M\le 0$. Then we can calculate $\langle \varphi_{j,k,l}|\varphi_{j,k,l}\rangle $ using  (\ref{CHexp}) and (\ref{differentiaalvgl}), yielding
\begin{eqnarray*}
\langle \varphi_{j,k,l}|\varphi_{j,k,l}\rangle&=&\langle (-R^2-\nabla^2+2\mE+M)\varphi_{j-1,k,l} |\varphi_{j,k,l} \rangle\\
&=&\langle \varphi_{j-1,k,l} |(-R^2-\nabla^2-2\mE-M)\varphi_{j,k,l} \rangle\\
&=&8j(2j+M+2k-2)\langle  \varphi_{j-1,k,l}|\varphi_{j-1,k,l}\rangle.
\end{eqnarray*}
Now in the case $M<0$ we can choose $k=0$ and $j$ small enough, but larger than zero, to make the factor $2j+M-2$ negative. This means that either $\langle \varphi_{j,0,1}|\varphi_{j,0,1}\rangle$ or $\langle  \varphi_{j-1,0,1}|\varphi_{j-1,0,1}\rangle$ is negative,  proving that the inner product is not positive definite. If $M=0$ we find that $\langle \varphi_{1,0,1}|\varphi_{1,0,1}\rangle=0$, proving again that the inner product is not positive definite.

If, on the other hand, $(R^2)^\dagger=-\nabla^2$ and $(\nabla^2)^\dagger=-R^2$, we know that $(2\mE+M)^\dagger=\frac{1}{2}([\nabla^2, R^2])^\dagger=(2\mE+M)$ and we obtain in a similar fashion
\[
\langle \varphi_{j,k,l}|\varphi_{j,k,l}\rangle=-8j(2j+M+2k-2)\langle  \varphi_{j-1,k,l}|\varphi_{j-1,k,l}\rangle.
\]
Now for the case where $j$ and $k$ are big enough we find that either $\langle \varphi_{j,k,l}|\varphi_{j,k,l}\rangle$ or $\langle \varphi_{j-1,k,l}|\varphi_{j-1,k,l}\rangle$ has to be negative.
 \end{proof}

\subsection{Mehler formula}

We are now able to establish a Mehler formula for the super spherical Hermite polynomials when $M>0$. 
We start from the basis $\{ \varphi_{j,k,l} \}$ considered in theorem \ref{orthsuperCH}. Normalizing this basis, according to formula (\ref{normCH}),  yields
\[
\phi_{j,k,l} =  \varphi_{j,k,l} / \sqrt{\gamma^M_{j,k}},
\]
which is an orthonormal basis with respect to $\langle\cdot|\cdot\rangle_2$. Using the integral expression for the inner product $\langle .|.\rangle_2$, the basis $\{\phi_{j,k,l}\}$ and the general fractional Fourier transform (see theorem \ref{fracfourthm}) we obtain that, formally,
\begin{eqnarray*}
&&\sum_{j,k,l} T(\phi_{j,k,l})(\bold{x}) \,e^{i \alpha (2j+k)} \,\phi_{j,k,l}(\bold{y})\\ &=& \left(\pi (1- e^{2i \alpha})\right)^{-M/2} \exp{\frac{ 4 e^{i \alpha} \langle \bold{x},\bold{y} \rangle - (1+ e^{2i \alpha})(R^2 + R_{\bold{y}}^2)}{2- 2e^{2i \alpha}}}.
\end{eqnarray*}
Using the explicit expression for the spherical Hermite polynomials in terms of Laguerre polynomials (see (\ref{CHLaguerre})) and the normalization (\ref{normCH}) we then find
\begin{eqnarray*}
&&\sum_{j,k,l} \frac{2j! e^{i \alpha (2j+k)}}{\Gamma(j+\frac{M}{2}+k)} L_j^{\frac{M}{2}+k-1}(R^2)L_j^{\frac{M}{2}+k-1}(R_{\bold{y}}^2)  T(H_{k}^{(l)})(\bold{x}) H_{k}^{(l)}(\bold{y}) \\
&=& \left(\pi (1- e^{2i \alpha})\right)^{-M/2} \exp{\frac{ 2 e^{i \alpha} \langle \bold{x},\bold{y} \rangle - e^{2i \alpha}(R^2 + R_{\bold{y}}^2)}{1- e^{2i \alpha}}}.
\end{eqnarray*}
This equation can be simplified again for $M>1$ with the explicit calculation of the summation $F_k(\bold{x},\bold{y})=\sum_lT({H}_k^l(\bold{x}))H_k^l(\bold{y})$. These functions $F_k$ were already constructed in \cite{DBS5} as a consequence of a Funk-Hecke theorem on the supersphere.

\begin{lemma}[Reproducing kernel]
\label{superrepr}
Let $M>1$. Then 
\[
G_{k}(\bold{x},\bold{y}) = \frac{2k+M-2}{M-2} \,\frac{\Gamma(M/2)}{2 \pi^{M/2}} \, C^{(M-2)/2}_k (\langle \bold{x},\bold{y} \rangle)
\]
with $C^{(M-2)/2}_k$ a Gegenbauer polynomial, is a reproducing kernel for the space $\cH_k$, i.e.
\[
\int_{SS,x} H_l(\bold{x}) G_{k}(\bold{x},\bold{y}) = \delta_{kl} H_l(\bold{y}) \mbox{ mod }(R_{\bold{y}}^2-1), \quad \mbox{for all } H_l \in \cH_l.
\]
\end{lemma}

\begin{proof}
See \cite{DBS5}, corollary 5. We have rewritten the Legendre polynomials used there in terms of Gegenbauer polynomials.
 \end{proof}

Making this function homogeneous, using equation (\ref{superint2}) and the fact that $\int_{SS}R^2P=\int_{SS}P$ yields
\begin{eqnarray*}
& &\int_{\mR^{m|2n}} H_l(\bold{x})\frac{2k+M-2}{M-2} \frac{\Gamma(M/2)}{2 \pi^{M/2}} (R^2R_{\bold{y}}^2)^{(k/2)} C^{(M-2)/2}_k \left(\frac{\langle \bold{x},\bold{y} \rangle}{(R^2R_{\bold{y}}^2)^{(1/2)}} \right) \exp (-R^2)\\
&=&\frac{\Gamma((k+l+M)/2)} {2}\int_{SS,x} H_l(\bold{x}) F_{k}(\bold{x},\frac{\bold{y}}{\sqrt{R_{\bold{y}}^2}})(R_{\bold{y}}^2)^{k/2} \\
&=&\delta_{kl}\frac{\Gamma(k+M/2)} {2}H_l(\bold{y}).
\end{eqnarray*}
Hence we conclude
\begin{eqnarray*}
F_k(\bold{x},\bold{y}) &=& \sum_{l=1}^{\dim \cH_{k}} T({H}_k^{(l)}(\bold{x}))H_k^{(l)}(\bold{y})\\ 
&=&\frac{2k+M-2}{M-2} \frac{\Gamma(M/2)}{2 \pi^{M/2}} (R^2R_{\bold{y}}^2)^{(k/2)} C^{(M-2)/2}_k \left(\frac{\langle \bold{x},\bold{y} \rangle}{(R^2R_{\bold{y}}^2)^{(1/2)}} \right)
\end{eqnarray*}
for a basis of $\cH_k$ as in equation (\ref{orthbasissuper}). Putting everything together then yields the following Mehler formula in superspace
\begin{eqnarray}
\nonumber
&&\sum_{j,k}\frac{2j! e^{i \alpha (2j+k)}}{\Gamma(j+\frac{M}{2}+k)}    L_j^{\frac{M}{2}+k-1}(R^2)L_j^{\frac{M}{2}+k-1}(R_{\bold{y}}^2) F_k (\bold{x},\bold{y}) 
\\
&=& \left(\pi (1- e^{2i \alpha})\right)^{-\frac{M}{2}} \exp{\frac{  2 e^{i \alpha} \langle \bold{x},\bold{y} \rangle - e^{2i \alpha}(R^2 + R_{\bold{y}}^2)}{1- e^{2i \alpha}}}.
\label{superMehler2}
\end{eqnarray}
So far, we have only established this formula formally. We now show that we have actual convergence.

\begin{theorem}
For $\ux,\uy\in\mR^m$, $\alpha\in\mR$ and $M>1$, the series in equation (\ref{superMehler2}) converges pointwise.
\end{theorem}

\begin{proof}
Both sides in equation \eqref{superMehler2} are functions of $\langle \bold{x},\bold{y}\rangle$, $R^2$ and $R_{\bold{y}}^2$. This equation can therefore be written as
\[
\sum_{j,k}g_{j,k}(\langle \bold{x},\bold{y}\rangle,R^2,R_{\bold{y}}^2)=g(\langle \bold{x},\bold{y}\rangle,R^2,R_{\bold{y}}^2).
\]
As the functions $g_{j,k}$ are polynomials in $\langle \bold{x},\bold{y}\rangle$, $R^2$ and $R_{\bold{y}}^2$, they can be written as a Taylor expansion in the anti-commuting variables
\begin{eqnarray*}
g_{j,k}(\langle \bold{x},\bold{y}\rangle,R^2,R_{\bold{y}}^2)&=&\sum_{p,q=0}^n\sum_{r=0}^{2n}\frac{\theta^{2p}\theta_{\uyb}^{2q}\langle \uxb,\uyb\rangle^r}{p!q!r!}\\
&&\times (\frac{\partial}{\partial a^2})^p(\frac{\partial}{\partial b^2})^q(\frac{\partial}{\partial c})^rg_{j,k}(c,a^2,b^2)
\end{eqnarray*}
with $a^2=r^2$, $b^2=r_{\uy}^2$ and $c=\langle\ux,\uy\rangle$. This also holds for
\begin{eqnarray*}
g(\langle \bold{x},\bold{y}\rangle,R^2,R_{\bold{y}}^2)&=&\left(\pi (1- e^{2i \alpha})\right)^{-\frac{M}{2}} \exp{\frac{  2 e^{i \alpha} \langle \bold{x},\bold{y} \rangle - e^{2i \alpha}(R^2 + R_{\bold{y}}^2)}{1- e^{2i \alpha}}}\\
&=&\left(\pi (1- e^{2i \alpha})\right)^{-\frac{M}{2}} \exp{\frac{  2 e^{i \alpha}c -e^{2i \alpha}(a^2 + b^2)}{1- e^{2i \alpha}}} \\
&&\times \exp{\frac{  2 e^{i \alpha} \langle \uxb , \uyb \rangle - e^{2i \alpha}(\theta^2 + \theta_{\uyb}^2)}{1- e^{2i \alpha}}}.
\end{eqnarray*}
From corollary \ref{OmMehler3} we find
\[
\sum_{j,k}g_{j,k}(c,a^2,b^2)=g(c,a^2,b^2).
\]
Remark \ref{Mehlerafleiden} implies that arbitrary derivatives with respect to $c$, $a^2$ and $b^2$ can be brought inside the summation with the convergence still holding. So we can take the Taylor series of both sides and the theorem is proven.
 \end{proof}

\section{Conclusions and summary}
\setcounter{equation}{0}
\label{conclSummary}

In this paper, we have given a detailed treatment of Hermite type polynomials related to three different symmetries: $O(m)$ (orthogonal symmetry), $\cG < O(m)$ (finite reflection group symmetry) and $O(m) \times Sp(2n)$ (superspace symmetry). In each case, it was possible to define two types of Hermite polynomials, namely cartesian Hermite polynomials and spherical Hermite polynomials. In the cases of $O(m)$, $\cG$ and $Sp(2n)$, both types of polynomials turned out to be orthogonal with respect to the canonical inner product. In the full superspace case of $O(m) \times Sp(2n)$, we obtained that the spherical Hermite polynomials are not orthogonal with respect to the canonical inner product. We presented a detailed analysis of this lack of orthogonality. Then we gave a construction of a new inner product, which restores the orthogonality of the spherical Hermite polynomials but destroys the orthogonality of the cartesian Hermite polynomials.

We have summarized all these results in two tables. In Table \ref{default}, we give an extensive overview of the different types of symmetry and compare their analogies and differences. For the superspace case, the purely fermionic ($Sp(2n)$) case is given in a separate column. In Table \ref{default2} we restrict ourselves to the full superspace case and give a comparison between the two inner products that we have considered. We give the adjoints of the relevant operators as well as the differences in orthogonality of the two types of Hermite polynomials.

As already mentioned, the inner product we have constructed in section \ref{innerprodsSuper} has only been defined for a weighted space of polynomials. It is possible to extend this inner product to broader function spaces, such as $\cS(\mR^{m})_{m|2n}$. As this is a technical matter requiring subtle estimates, we postpone this to a subsequent paper (see \cite{CDBHilbert}).

The results obtained in this paper allow to study several other interesting problems in the future.

First of all, as we now have a new inner product on superspace that makes hamiltonians of the type
\[
H = -\frac{1}{2}\nabla^2 + V(R^{2})
\]
self-adjoint, we can make a thorough study of such systems and the related Schr\"odinger equations. This should allow to shed new light on the results obtained in e.g. \cite{MR967935,ZHANG} and to put them together in one theory.

In \cite{DBS3} we also introduced a second class of polynomials in superspace, namely the spherical Gegenbauer polynomials. It is expected that the results obtained here in combination with the new view on supersphere integration given in \cite{CDBS1} will allow to obtain orthogonality properties of these polynomials on the superball.

Next, the obtained Mehler formulas in superspace should allow to study the radial behavior of the super Fourier transform and its relation  with the classical Hankel transform. More generally, we have now the tools available to make a complete study in superspace of the holomorphic semigroup 
\[
e^{z (-\nabla^2 +R^{2})}, \quad z \in \mC, \Re z \geq 0
\]
in the sense of \cite{Folland,Howe}. Note that the choice $z= i \pi/4$ leads to the super Fourier transform (see formula (\ref{FTSuperExp})).

Another important direction for further research is in the context of radial deformations. The study of such deformations has recently arisen as a new and exciting topic in harmonic analysis. In the orthogonal situation, one special radial deformation has been studied in \cite{MR2134314,MR2401813}. The Dunkl case has been considered in \cite{Orsted2}. There, the authors introduce a radial deformation parameter in the $\mathfrak{sl}_{2}$ relations satisfied by the Dunkl Laplacian and make a detailed study of the related analysis. Very recently, an even more general radial deformation in the context of Dirac operators has been realized in \cite{DBOrsted}. 
It is expected that also the superspace representation of $\mathfrak{sl}_{2}$ (as well as its Dirac counterpart given by $\mathfrak{osp}(1|2)$) can be radially deformed, i.e. that it would be possible to replace the super Laplace operator $\nabla^2$ and $R^{2}$ by radially deformed operators in such a way that the $\mathfrak{sl}_{2}$ relations are preserved. It would be very interesting to see to what extent the theory of radial deformations can be established in the setting of superspaces and whether the Hermite polynomials related to these new deformations have similar orthogonality properties as established in this paper.

Finally, Table \ref{default} suggests that there is a type of symmetry missing in the current scheme. This is the case of a realization of 
 $\mathfrak{sl}_{2}$ which is only invariant under a (finite) subgroup of $Sp(2n)$, thus establishing the symplectic counterpart of the theory of Dunkl operators. It is at this point not entirely clear whether such a deformation is feasible in all generality in the framework of superspaces, but it would give a very satisfying unifying picture.
Note that in the special case of a superspace with $2n$ commuting and $2n$ anti-commuting variables, so with invariance $O(2n) \times Sp(2n)$, one has already established analogs of various Calogero-Sutherland systems (see e.g. \cite{MR1070940, MR1608453, MR2025382}). Although the hamiltonians considered in those papers don't contain the fermionic Laplace operator $\nabla^{2}_{f}$, contrary to formula (\ref{SchrodEq}), this still hints at possible generalizations to the non-supersymmetric case.

\begin{sidewaystable}[htdp]
\caption{Summary - 4 types of symmetry}
\begin{center}
\scalebox{0.75}{\begin{tabular}{c||c|c|c|c}
Symmetry&$O(m)$& $\cG < O(m)$& $Sp(2n)$&$O(m) \times Sp(2n)$\\ \vspace{-3mm} &&&& \\
&orthogonal&finite reflection group& symplectic& full superspace\\ \vspace{-3mm} &&&& \\ \hline \hline \vspace{-3mm} &&&& \\
Basic function space& $L_{2}(\mR^{m})$&$L_{2}(\mR^{m}, w_{\kappa}(\ux) dV(\ux))$& $\Lambda_{2n}$ & $L_{2}(\mR^{m})\otimes \Lambda_{2n}$\\ \vspace{-3mm} &&&& \\
Generators of $\mathfrak{sl}_{2}$& $\nabla^2_{b}$&$\Delta_{\kappa}$&$\nabla^2_{f}$& $\nabla^2 = \nabla^2_{b} + \nabla^2_{f}$\\ \vspace{-3mm} &&&& \\
&$r^{2}$&$r^{2}$&$\theta^{2}$&$R^{2} = \theta^{2}+r^{2}$\\ \vspace{-3mm} &&&& \\
& $\mE_{b} + \frac{m}{2}$& $\mE_{b} + \frac{\mu}{2}$ & $\mE_{f} -n $&$\mE + \frac{m-2n}{2}$\\ \vspace{-3mm} &&&& \\
Dimension& $m$&$\mu = m + 2\sum_{\alpha \in R_+} \kappa_{\alpha}$&$-2n$&$M = m -2n$\\ \vspace{-3mm} &&&& \\
Spaces of harmonics & $\cH_{k}^{b} =\ker{\nabla^2_{b}} \cap Pol_{k}$ &$\cH_{k}^{\cD} =\ker{\Delta_{\kappa}} \cap Pol_{k}$ & $\cH_{k}^{f} =\ker{\nabla^2_{f}} \cap \Lambda_{2n}^{k}$ & $\cH_{k} =\ker{\nabla^2} \cap \cP_{k}$\\ \vspace{-3mm} &&&& \\
Reproducing kernel & $C^{(m-2)/2}_{k}$ & $V_{\kappa}(C^{(\mu-2)/2}_{k})$& $C^{(-n-1)}_{k}$& $C^{(M-2)/2}_{k}$\\ \vspace{-3mm} &&&& \\ \hline \vspace{-3mm} &&&& \\
Related quantum system& $\frac{1}{2}(-\nabla^2_{b}  + r^{2}) \psi = E \psi
$ & $\frac{1}{2}(-\Delta_{\kappa}  + r^{2}) \psi = E \psi
$& $\frac{1}{2}(-\nabla^2_{f}+\theta^2) \psi = E \psi $ & $\frac{1}{2}(-\nabla^2+R^2) \psi = E \psi $\\ \vspace{-3mm} &&&& \\
& PDE&PDE + difference terms& system of algebraic equations& system of PDEs\\ \vspace{-3mm} &&&& \\
Fourier transform & $e^{ \frac{i \pi m}{4}} e^{\frac{i \pi}{4}(\nabla^2_{b} - r^{2})}$
&$e^{ \frac{i \pi \mu}{4}} e^{\frac{i \pi}{4}(\Delta_{\kappa} - r^{2})}$&$e^{-\frac{i \pi n}{2} } e^{ \frac{i \pi }{4}(\nabla^2_{f} -\theta^2)}$ &$e^{\frac{ i \pi M}{4} } e^{\frac{i \pi}{4}(\nabla^2 -R^2)}$ \\ \vspace{-3mm} &&&& \\
&$(2 \pi)^{-\frac{m}{2}} \int_{\mR^{m}} e^{-i\langle \ux,\uy\rangle} f(\ux) dV(\ux)$&$ c_{\kappa}^{-1} \int_{\mR^m} D(\ux,-i\uy) f(\ux) w_{\kappa}(\ux) dV(\ux)$& $(2 \pi)^{n}  \int_{B,x} \exp{(- i \langle \uxb , \uyb \rangle)}f(\bold{x})$& $(2 \pi)^{-\frac{M}{2}} \int_{\mR^{m|2n}} \exp{(- i \langle \bold{x},\bold{y} \rangle)}f(\bold{x})$\\ \vspace{-3mm} &&&& \\
&$\langle \ux,\uy\rangle = \sum_{i=1}^{m}x_{i}y_{i}$& $D(\ux,-i\uy)$ in general unknown&$\langle \uxb,\uyb\rangle = -\frac{1}{2} \sum_{j=1}^{n}({x \grave{}}_{2j-1}{y \grave{}}_{2j} - {x \grave{}}_{2j} {y \grave{}}_{2j-1})$ &$\langle \bold{x},\bold{y} \rangle = \langle \ux,\uy\rangle + \langle \uxb,\uyb\rangle  $\\ \vspace{-3mm} &&&& \\
Integration & Lebesgue integral&weighted Lebesgue integral& Berezin integral $\int_{B}$& $\int_{\mR^{m|2n}} = \int_{B} \int_{\mR^{m}}dV(\ux)$ \\ \vspace{-3mm} &&&& \\ \hline \vspace{-3mm} &&&& \\
Cartesian Hermite functions & $\psi_{k_{1}, \ldots, k_{m}}^{b}$ & $\psi_\nu^{\cD}$ & $\psi_{ l_{1}, \ldots, l_{2n}}^{f}$& $\psi_{k_{1}, \ldots, k_{m}; l_{1}, \ldots, l_{2n}}$\\ \vspace{-3mm} &&&& \\
Energy & $\frac{m}{2} + \sum_{i=1}^{m}k_{i}$& $\frac{\mu}{2} + |\nu|$& $-n + \sum_{i=1}^{2n}l_{i}$& $\frac{M}{2} + \sum_{i=1}^{m}k_{i} + \sum_{i=1}^{2n}l_{i}$ \\ \vspace{-3mm} &&&& \\
Spherical Hermite functions & $\phi_{j,k,l}^{b}$ & $\phi_{j,k,l}^{\cD}$& $\phi_{j,k,l}^{f}$ & $\phi_{j,k,l}$ \\ \vspace{-3mm} &&&& \\
Energy & $\frac{m}{2} + (2j + k)$&$\frac{\mu}{2} + (2j + k)$&$-n + (2j + k)$&$\frac{M}{2} + (2j + k)$ \\ \vspace{-3mm} &&&& \\ \hline \vspace{-3mm} &&&& \\
Canonical inner product & $\langle f, g \rangle_{L_{2}} = \int_{\mR^{m}} f \overline{g} dV(\ux)$ &$\langle f, g \rangle_{L_{2}} = \int_{\mR^{m}} f \overline{g} w_{\kappa}(\ux)dV(\ux)$& $\langle f|g \rangle_{\Lambda_{2n}} = \int_{B} f (\ast \overline{g})$& $\langle f|g \rangle_{1} =\int_{\mR^{m|2n}} f (* \overline{g}) $ \\ \vspace{-3mm} &&&& \\ 
Orth. cartesian Hermites &$\langle \psi_{k_{1}, \ldots, k_{m}}^{b} ,\psi_{l_{1}, \ldots, l_{m}}^{b} \rangle_{L_{2}}$ & $\langle \psi_\nu^{\cD} ,\psi_\mu^{\cD} \rangle_{L_{2}}$&$\langle \psi_{l_{1}, \ldots, l_{2n}}^{f}, \psi_{q_{1}, \ldots, q_{2n}}^{f}\rangle_{\Lambda_{2n}}$&$\langle \psi_{k_{1}, \ldots, k_{m}; l_{1}, \ldots, l_{2n}}, \psi_{p_{1}, \ldots, p_{m}; q_{1}, \ldots, q_{2n}}\rangle_{1}$\\ \vspace{-3mm} &&&& \\
& $  = \delta_{k_{1} l_{1}} \ldots \delta_{k_{m} l_{m}}$& $= \delta_{\mu \nu}$&$ =  \delta_{l_{1} q_{1}} \ldots \delta_{l_{2n} q_{2n}}$&$ = \delta_{k_{1} p_{1}} \ldots \delta_{k_{m} p_{m}} \delta_{l_{1} q_{1}} \ldots \delta_{l_{2n} q_{2n}}$\\ \vspace{-3mm} &&&& \\ \hline \vspace{-3mm} &&&& \\
Orth. spherical Hermites &$\langle \phi_{j_{1},k_{1},l_{1}}^{b} , \phi_{j_{2},k_{2},l_{2}}^{b}\rangle_{L_{2}}$& $\langle \phi_{j_{1},k_{1},l_{1}}^{\cD} , \phi_{j_{2},k_{2},l_{2}}^{\cD}\rangle_{L_{2}} $& $\langle \phi_{j_{1},k_{1},l_{1}}^{f} , \phi_{j_{2},k_{2},l_{2}}^{f}\rangle_{\Lambda_{2n}}$& $\langle \phi_{j_{1},k_{1},l_{1}} , \phi_{j_{2},k_{2},l_{2}}\rangle_{1} $\\ \vspace{-3mm} &&&& \\
&$ = \delta_{j_{1} j_{2}} \delta_{k_{1} k_{2}} \delta_{l_{1} l_{2}}$& $= \delta_{j_{1} j_{2}} \delta_{k_{1} k_{2}} \delta_{l_{1} l_{2}}$& $= \delta_{j_{1} j_{2}} \delta_{k_{1} k_{2}} \delta_{l_{1} l_{2}}$&${\bf\neq} \delta_{j_{1} j_{2}} \delta_{k_{1} k_{2}} \delta_{l_{1} l_{2}}$\\ \vspace{-3mm} &&&& \\
\end{tabular}}
\end{center}
\label{default}
\end{sidewaystable}%

\begin{sidewaystable}[htdp]
\caption{Inner products in the full superspace case ($M >0$)}
\begin{center}
\begin{tabular}{c||c|c}
Inner product & $\langle f|g \rangle_{1} =\int_{\mR^{m|2n}} f (* \overline{g}) $& $\langle f|g \rangle_{2} =\int_{\mR^{m|2n}} f \overline{T(g)} $\\
\vspace{-3mm}&&\\ \hline \hline
\vspace{-3mm}&&\\
Function space & $L_{2}(\mR^{m})\otimes \Lambda_{2n}$ & $\cP \exp(-R^{2}/2)$ \\
\vspace{-3mm}&&\\ \hline
\vspace{-3mm}&&\\
Adjoints: &&\\ \vspace{-3mm}&&\\
$(a_{i}^{\pm})^{\dagger}$ &$a_{i}^{\mp}$ & ? \\\vspace{-3mm}&&\\
$(b_{i}^{\pm})^{\dagger}$ &$b_{i}^{\mp}$ & ? \\\vspace{-3mm}&&\\
$(\nabla^2)^{\dagger} $& $\nabla^2_{b}-\theta^{2}$ & $\nabla^2$ \\\vspace{-3mm}&&\\
$(R^{2})^{\dagger} $& $r^{2} -\nabla^2_{f}$ & $R^{2}$ \\\vspace{-3mm}&&\\
$(\mE + \frac{M}{2})^{\dagger} $& $\mE_{f} -\mE_{b} -\frac{m}{2} -n$ &$-(\mE + \frac{M}{2})$\\\vspace{-3mm}&&\\
$H^{\dagger} = \frac{1}{2}(-\nabla^2+R^2)^{\dagger} $ &$\frac{1}{2}(-\nabla^2+R^2)$ & $\frac{1}{2}(-\nabla^2+R^2)$ \\ \vspace{-3mm}&&\\ \vspace{-3mm}&&\\ \hline\vspace{-3mm}&&\\
Orth. cartesian Hermites & $\langle \psi_{k_{1}, \ldots, k_{m}; l_{1}, \ldots, l_{2n}}, \psi_{p_{1}, \ldots, p_{m}; q_{1}, \ldots, q_{2n}}\rangle_{1}$ & $\langle \psi_{k_{1}, \ldots, k_{m}; l_{1}, \ldots, l_{2n}}, \psi_{p_{1}, \ldots, p_{m}; q_{1}, \ldots, q_{2n}}\rangle_{2}$\\
&$ = \delta_{k_{1} p_{1}} \ldots \delta_{k_{m} p_{m}} \delta_{l_{1} q_{1}} \ldots \delta_{l_{2n} q_{2n}}$&$ \neq \delta_{k_{1} p_{1}} \ldots \delta_{k_{m} p_{m}} \delta_{l_{1} q_{1}} \ldots \delta_{l_{2n} q_{2n}}$\\ \vspace{-3mm}&&\\ \hline \vspace{-3mm}&&\\
Orth. spherical Hermites &$\langle \phi_{j_{1},k_{1},l_{1}} , \phi_{j_{2},k_{2},l_{2}}\rangle_{1} $ &$\langle \phi_{j_{1},k_{1},l_{1}} , \phi_{j_{2},k_{2},l_{2}}\rangle_{2} $ \\
&${\bf\neq} \delta_{j_{1} j_{2}} \delta_{k_{1} k_{2}} \delta_{l_{1} l_{2}}$& $ = \delta_{j_{1} j_{2}} \delta_{k_{1} k_{2}} \delta_{l_{1} l_{2}}$\\
\end{tabular}
\end{center}
\label{default2}
\end{sidewaystable}

\section{Appendix}
\setcounter{equation}{0}

The Hermite polynomials $H_k$ for $k\in\mN$ are defined by their Rodrigues formula as $H_{k}(t) = (-1)^{k} \exp{(t^2)} \frac{d^k}{dt^k} \exp{(-t^2)}$ and are given explicitly by
\begin{eqnarray*}
H_k(t)&=&\sum_{j=0}^{\lfloor k/2\rfloor}(-1)^j\frac{2^{k-2j}k!}{(k-2j)!j!}t^{k-2j}.
\end{eqnarray*}
They satisfy the orthogonality relation
\begin{eqnarray*}
\int_{-\infty}^\infty H_k(t)H_l(t)\exp(-t^2)dt&=&\delta_{kl}k!2^k\sqrt{\pi}.
\end{eqnarray*}

The generalized Laguerre polynomials $L_k^{(\alpha)}$ for $k\in\mN$  are defined as
\begin{eqnarray}
L_k^{(\alpha)}(t)&=&\sum_{j=0}^{k}\frac{\Gamma(k+\alpha+1)}{j!(k-j)!\Gamma(j+\alpha+1)}(-t)^j
\label{genLagdef}
\end{eqnarray}
and satisfy the orthogonality relation (when $\alpha>-1$)
\begin{eqnarray*}
\int_{0}^\infty t^{\alpha} L^{(\alpha)}_k(t)L^{(\alpha)}_l(t)\exp(-t)dt&=&\delta_{kl}\frac{\Gamma(k+\alpha+1)}{k!}.
\end{eqnarray*}
The Hermite polynomials can be expressed in terms of the generalized Laguerre polynomials by
\begin{eqnarray*}
H_{2k}(t)=(-1)^k2^{2k}k!L_k^{(-\frac{1}{2})}(t^2) &\mbox{and}&H_{2k+1}(t)=(-1)^k2^{2k+1}k!tL_k^{(\frac{1}{2})}(t^2).
\end{eqnarray*}

The Gegenbauer polynomials $C^{(\alpha)}_k(t)$ are a special case of the Jacobi polynomials. For $k\in\mN$ and $\alpha>-1/2$ they are defined as
\begin{eqnarray}
C_k^{(\alpha)}(t)&=&\sum_{j=0}^{\lfloor k/2\rfloor}(-1)^j\frac{\Gamma(k-j+\alpha)}{\Gamma(\alpha)j!(k-2j)!}(2t)^{k-2j}
\label{GegenbauerCoeffs}
\end{eqnarray}
and satisfy the orthogonality relation
\begin{eqnarray*}
\int_{-1}^1C_k^{(\alpha)}(t)C_l^{(\alpha)}(t)(1-t^2)^{\alpha-\frac{1}{2}}dt&=&\delta_{kl}\frac{\pi2^{1-2\alpha}\Gamma(k+2\alpha)}{k!(k+\alpha)(\Gamma(\alpha))^2}.
\end{eqnarray*}

\end{document}